\documentclass[3p]{elsarticle} 

\usepackage{amssymb}
\usepackage{amsmath}

\usepackage[ruled]{algorithm2e}
\usepackage{amsfonts}
\usepackage{mathtools}
\usepackage{amsmath,bm}
\usepackage{amsthm}
\usepackage{graphicx}
\usepackage{mathrsfs}
\usepackage{float}

\usepackage[colorlinks=true, allcolors=blue]{hyperref}
\newtheorem{proposition}{Proposition}

\begin{document}

\begin{frontmatter}





\title{An adjoint method for training data-driven reduced-order models}

\author[label1,label2]{Donglin Liu\corref{cor1}\fnref{eq}}
\ead{donglin.liu@med.lu.se}
\author[label1]{Francisco García Atienza\fnref{eq}}
\ead{fragarat@gmail.com}
\author[label1]{Mengwu Guo}
\ead{mengwu.guo@math.lu.se}
\affiliation[label1]{organization={Centre for Mathematical Sciences, Lund Univeristy}, country={Sweden}}
\affiliation[label2]{organization={Department of Experimental Medical Science, Lund Univeristy}, country={Sweden}}

\cortext[cor1]{Corresponding author.}
\fntext[eq]{The first two authors contributed equally to this work.}
            
\begin{abstract}
Reduced-order modeling lies at the interface of numerical analysis and data-driven scientific computing, providing principled ways to compress high-fidelity simulations in science and engineering. We propose a training framework that couples a continuous-time form of operator inference with the adjoint-state method to obtain robust data-driven reduced-order models. This method minimizes a trajectory-based loss between reduced-order solutions and projected snapshot data, which removes the need to estimate time derivatives from noisy measurements and provides intrinsic temporal regularization through time integration. We derive the corresponding continuous adjoint equations to compute gradients efficiently and implement a gradient-based optimizer to update the reduced model parameters. Each iteration only requires one forward reduced-order solve and one adjoint solve, followed by inexpensive gradient assembly, making the method attractive for large-scale simulations.
We validate the proposed method on three partial differential equations: viscous Burgers’ equation, the two-dimensional Fisher-KPP equation, and an advection-diffusion equation. We perform systematic comparisons against standard operator inference under two perturbation regimes, namely reduced temporal snapshot density and additive Gaussian noise. For clean data, both approaches deliver similar accuracy, but in situations with sparse sampling and noise, the proposed adjoint-based training provides better accuracy and enhanced roll-out stability.
\end{abstract}



\begin{keyword}
reduced-order modeling \sep operator inference \sep constrained optimization \sep adjoint method \sep scientific machine learning



\end{keyword}

\end{frontmatter}



\section{Introduction}
High-fidelity numerical simulation is central to digital twins, which link physical assets and virtual replicas through a bidirectional exchange of information: measurement data calibrate and update the virtual model, while model predictions guide monitoring, control, and decision-making for the physical system \cite{National}. For deployment, a digital twin must deliver predictions that are both fast and reliable, yet faithfully resolving the underlying physics often requires high-dimensional dynamical models with prohibitive computational costs. This challenge is especially acute in safety-critical settings, where computational efficiency must be paired with credibility and explainability of model-based decisions \cite{Gunning}.

Reduced-order models (ROMs) provide interpretable computational surrogates that compress high-fidelity, high-dimensional simulations into predominant low dimensionality, enabling efficient computation while preserving essential physical behaviors \cite{benner2015survey, ghattas2021learning}. Classical projection-based ROMs construct reduced dynamics intrusively by projecting a full-order model (FOM) onto a low-dimensional subspace \cite{benner2015survey,berkooz1993proper,quarteroni2015reduced}. In contrast, non-intrusive, data-driven ROMs avoid direct interaction with the FOM implementation and instead learn low-dimensional latent dynamics primarily from simulation data, which reduces implementation burden and extends model reduction to legacy codes and black-box simulators.  
A broad range of data-driven numerical methods falls under this umbrella, including dynamic mode decomposition (DMD) \cite{Steven2015,annurev015835,doi22M1481658,104913868}, operator inference (OpInf) \cite{ghattas2021learning, kramer2024learning, PEHERSTORFER2016196, mcquarrie2025bayesian}, the Loewner framework \cite{doi130914619}, interpolation-based methods \cite{Franz16032014, doiJ065798},  Gaussian process surrogate models \cite{GUO2018807,CICCI20231}, and deep learning approaches \cite{fresca2022pod, articleManzoni}. 
In this work, we focus on OpInf, which identifies structured low-dimensional operators directly from trajectory data while enforcing a prescribed model form.

A key practical limitation of data-driven ROMs is the quantity (e.g., sparsely sampled measurements) and quality (e.g., noisy observations) of available training data. Standard OpInf commonly identifies reduced operators by regressing against estimated time derivatives of state snapshots, which makes the method sensitive to both measurement noise and coarse temporal sampling. In particular, explicit numerical differentiation of state data amplifies noise and biases the inferred operators under data scarcity, which typically destabilizes the time integration of nonlinear reduced-order systems. These issues motivate training formulations that avoid such numerical differentiation and incorporate temporal regularization through time integration, hence improving robustness while remaining computationally viable for complex systems. Related development in differentiable programming for dynamics learning \cite{sapienza2024differentiable}, such as neural ordinary differential equations (neural ODEs), shows that trajectory-based objectives can be optimized efficiently by differentiating through ODE solvers via adjoint methods \cite{chen2018neural,Antil2018,bradley2024pde}.

This work develops a continuous-time, gradient-based training framework for OpInf and computes gradients via the adjoint method. We minimize an integral trajectory misfit between the ROM solution and projected reduced-state data, thereby avoiding explicit time-derivative estimation and enabling intrinsic regularization through time integration. Gradients of this loss with respect to ROM parameters are evaluated by solving an adjoint-state system backward in time. Thus, each iterative step in the optimization process only requires one forward reduced-order solve and one adjoint solve, followed by additional algebra to assemble parameter gradients. In this way, our method mirrors the efficient optimization mechanism of neural ODEs while retaining the interpretability, robustness, and physical consistency of OpInf. We validate the proposed method on three canonical partial differential equations (PDEs): viscous Burgers' equation, the two-dimensional Fisher-KPP reaction-diffusion equation, and an advection-diffusion equation. For each numerical example, we construct a POD basis from full-order snapshots, initialize reduced operators using standard OpInf, and then refine them using the proposed adjoint-based training.  In all three examples, we compare standard OpInf and our proposed adjoint-based version under two perturbations: uniform temporal subsampling for reduced snapshot density, and additive Gaussian noise at multiple levels. For clean data, the proposed method and standard OpInf achieve comparable accuracy. However, when dealing with sparse sampling and noisy data, the adjoint-based training offers improved accuracy and greater stability during time integration.

The remainder of this paper is organized as follows. 
Section 2 reviews reduced-order modeling and OpInf. In Section 3, thereafter, we introduce the proposed trajectory-fitting scheme of OpInf using adjoint-state equations and discuss its computational implementation. Results for numerical examples are presented and discussed in Section 4, and concluding remarks are eventually made in Section 5.

\section{Reduced-order models via operator inference}
We consider a high–dimensional FOM
\begin{equation}
  \dot{\mathbf u}(t)=\mathbf F\big(\mathbf u(t),\mathbf s(t)\big), \qquad t\in[0,T],
  \label{eq:fom}
\end{equation}
with state $\mathbf u(t)\in\mathbb R^{n}$ and input $\mathbf s(t)\in\mathbb R^{m}$. 
In this work, we focus on a common quadratic structure \cite{PEHERSTORFER2016196}
\begin{equation}
  \dot{\mathbf u}(t)
  = \bm{\mathcal C}
  + \bm{\mathcal A}\,\mathbf u(t)
  + \bm{\mathcal H}\big(\mathbf u(t)\otimes\mathbf u(t)\big)
  + \bm{\mathcal B}\,\mathbf s(t), 
  \label{eq:fompoly}
\end{equation}
where $\bm{\mathcal C}\in\mathbb R^{n}$, $\bm{\mathcal A}\in\mathbb R^{n\times n}$,
$\bm{\mathcal H}\in\mathbb R^{n\times n^{2}}$, $\bm{\mathcal B}\in\mathbb R^{n\times m}$, and $\otimes$ denotes the Kronecker product \cite{LOAN200085}.

The reduced state $\mathbf q(t)\in\mathbb R^r$ approximates the FOM trajectory in an $r$–dimensional subspace with $r\ll n$,
\begin{equation*}
\mathbf u(t)\approx \mathbf V_r\mathbf q(t),
\end{equation*}
where $\mathbf V_r\in\mathbb R^{n\times r}$ collects a reduced basis by the proper orthogonal decomposition (POD), built from state snapshots
$\mathbf U=\big[\mathbf u(t_1)\ \cdots\ \mathbf u(t_k)\big]\in\mathbb R^{n\times k}$ collected at time instances $t_1,\cdots, t_k$.
In particular, let the singular value decomposition (SVD) of the snapshot matrix $\mathbf U$ be $\mathbf U=\bm\Phi\,\bm\Sigma\,\bm\Psi^\top$, where $\bm\Sigma$ is a diagonal matrix collecting the
singular values $\sigma_1\ge \cdots \ge \sigma_{\text{rank}(\mathbf U)}\ge 0$.
The rank-\(r\) POD basis is \(\mathbf V_r=\bm\Phi_{:,1:r}\), and we select $r<\text{rank($\mathbf U$)}$ by the cumulative–energy criterion
$\sum_{i=1}^r \sigma_i^2/\sum_{i=1}^{\text{rank}(\mathbf U)} \sigma_i^2\geq 1-\epsilon_r$, where $\epsilon_r\ll1$ is a prescribed tolerance. 
By the Eckart–Young–Mirsky theorem \cite{Eckart1936}, the POD projection is optimal in the least–squares sense and
\[
\sum_{i=1}^k \big\|\mathbf u(t_i)-\mathbf V_r\mathbf V_r^\top \mathbf u(t_i)\big\|_2^2
=\big\|\mathbf U-\mathbf V_r\mathbf V_r^\top \mathbf U\big\|_F^2
=\sum_{i=r+1}^{\text{rank}(\mathbf U)} \sigma_i^2.
\]

Projecting system \eqref{eq:fom} onto the subspace $\text{Col}(\mathbf V_r) \subset \mathbb{R}^n$ gives the reduced dynamics
\begin{equation*}
    \dot{\mathbf{q}}(t) =
    \mathbf{V}_r^{\top}\mathbf{F}( \mathbf{V}_r\mathbf{q}(t), \mathbf{s}(t)).
\end{equation*}
For the quadratic system \eqref{eq:fompoly}, the ROM preserves the polynomial form
\begin{align}
\dot{\mathbf q}(t)
&= \mathbf c+\mathbf A\mathbf q(t)+\mathbf H\big(\mathbf q(t)\otimes\mathbf q(t)\big)+\mathbf B\mathbf s(t)
\eqqcolon \mathbf f\big(\mathbf q(t),\mathbf s(t);\bm\theta\big),
\label{eq:redopinf}
\end{align}
with reduced operators 
$\mathbf c=\mathbf V_r^{\!\top}\bm{\mathcal C}\in\mathbb R^{r}$,
$\mathbf A=\mathbf V_r^{\!\top}\bm{\mathcal A}\mathbf V_r\in\mathbb R^{r\times r}$,
$\mathbf H=\mathbf V_r^{\!\top}\bm{\mathcal H}\,(\mathbf V_r\otimes\mathbf V_r)\in\mathbb R^{r\times r^2}$, and
$\mathbf B=\mathbf V_r^{\!\top}\bm{\mathcal B}\in\mathbb R^{r\times m}$.
We collect the parameters as $\bm\theta=[\mathbf c,\mathbf A,\mathbf H,\mathbf B]\in\mathbb{R}^{d}$ with $d=r+r^2+r^3+rm$ denoting the dimension of the parameter space (equivalently, $\operatorname{vec}(\bm\theta)\in\mathbb{R}^{d}$ by stacking $\mathbf c,\mathbf A,\mathbf H,\mathbf B$). When the high-fidelity operators $(\bm{\mathcal C},\bm{\mathcal A},\bm{\mathcal H},\bm{\mathcal B})$ are available, one can assemble $(\mathbf c,\mathbf A,\mathbf H,\mathbf B)$ intrusively via projection. In many applications, however, the solver is proprietary or a black box, demanding non-intrusive, data-driven approaches that learn reduced operators from projected state data $(\mathbf q,\dot{\mathbf q})$ and the input data $\mathbf s$.

Operator inference (OpInf) 
fits the ROM \eqref{eq:redopinf} directly in POD coordinates by linear least squares \cite{PEHERSTORFER2016196}. Given $\mathbf V_r$, we form reduced snapshots $\mathbf q(t_i)=\mathbf V_r^\top\mathbf u(t_i)$ and estimate $\dot{\mathbf q}(t_i)$ $(1\leq i\leq k)$ by finite differences, specifically using 2nd- and 6th-order stencils in this work. The regression problem is then
\begin{equation}
\min_{\bm\theta}
\sum_{i=1}^k
\Big\|
\mathbf c+\mathbf A\mathbf q(t_i)
+\mathbf H\big(\mathbf q(t_i)\otimes\mathbf q(t_i)\big)
+\mathbf B\mathbf s(t_i)
-\dot{\mathbf q}(t_i)
\Big\|_2^2,
\label{eq:lst_opinf}
\end{equation}
optionally with Tikhonov (ridge) regularization \cite{Willoughby, f2890069d4} or solved via truncated SVD (TSVD) of the snapshots' Gram matrix \cite{Hansen1987, doi:10.1137/090771806} to improve conditioning. OpInf is non-intrusive and preserves polynomial structures of the governing ODEs, but its accuracy hinges on reliable $\dot{\mathbf q}$ estimates. High-order stencils reduce truncation error yet can amplify noise. This sensitivity motivates the continuous-time, adjoint-based training proposed in the next Section, which avoids explicit time-derivative estimation.

\section{The adjoint method for operator inference}

\subsection{Continuous-time loss functional}

To avoid explicit differentiation as in the standard OpInf approach, we introduce a continuous-time loss functional that compares a trajectory reconstructed by integrating the learned operators against the observed data in an $L_2$ sense.
Let $\mathcal T=[0,T]$ and $\mathbf q(\cdot)\in\mathcal C^1(\mathcal T;\mathbb R^r)$ denote the ROM state function over time. 
In practice, we evaluate snapshots $\{(t_i,\mathbf q_{\mathrm{true}}(t_i))\}_{i=1}^k$ for the reduced states as discrete trajectory data. 
The continuous-time trajectory loss in the reduced space is then defined by
\begin{equation}
  \ell:\ \mathcal C^1(\mathcal T;\mathbb R^r)\to\mathbb R,\qquad
  \mathbf q(\cdot)\ \mapsto\ \int_0^T \big\|\mathbf q(t)-\mathbf q_{\mathrm{true}}(t)\big\|_2^2\,\mathrm{d}t=\int_0^T g(\mathbf{q}(t),t)\mathrm{d}t,
  \label{eq:continuous_loss}
\end{equation}
where $g(\mathbf{q}(t),t)=\big\|\mathbf q(t)-\mathbf q_{\mathrm{true}}(t)\big\|_2^2$, and the continuous trajectory $\mathbf q_{\text{true}}$ is approximated by temporal interpolation of the reduced snapshots.

We parametrize the ROM as in \eqref{eq:redopinf} and collect the reduced operators in 
$\bm\theta=[\,\mathbf c,\mathbf A,\mathbf H,\mathbf B\,]\in\mathbb R^d$. 
Given an input function $\mathbf s(t)$, the predicted trajectory $\tilde{\mathbf q}(\cdot;\bm\theta)$ is the solution of
\begin{equation*}
  \tilde{\mathbf q}(t;\bm\theta)
  = \tilde{\mathbf q}_0 
    + \int_0^t \!\Big[\mathbf c + \mathbf A\,\tilde{\mathbf q}(\tau;\bm\theta)
      + \mathbf H\!\big(\tilde{\mathbf q}(\tau;\bm\theta)\otimes \tilde{\mathbf q}(\tau;\bm\theta)\big)
      + \mathbf B\,\mathbf s(\tau)\Big]\,\mathrm d\tau .
\end{equation*}
Composing Eq.~\eqref{eq:continuous_loss} with the solution map $\bm\theta\mapsto\tilde{\mathbf q}(\cdot;\bm\theta)$ yields a reduced loss function with the parameters $\bm\theta$ as its variables:
\begin{equation}
  \tilde{\ell}(\bm\theta)
  := \ell\big(\tilde{\mathbf q}(\cdot;\bm\theta)\big)
  = \int_0^T \big\|\tilde{\mathbf q}(t;\bm\theta)-\mathbf q_{\mathrm{true}}(t)\big\|_2^2\,\mathrm dt .
  \label{eq:continuous_reduced_loss}
\end{equation}
The learning problem for determining the reduced operators $\bm\theta$ is then an unconstrained optimization 
\begin{equation*}
  \bm\theta^*\ \in\ \arg\min_{\bm\theta\in\mathbb R^d}\ \tilde{\ell}(\bm\theta).
\end{equation*}
To minimize the loss function $\tilde{\ell}(\bm\theta)$, we compute the total gradient with respect to the parameters. By the chain rule we have
\begin{equation*}
  \frac{\mathrm{d}\tilde{\ell}(\bm{\theta})}{\mathrm{d}\bm{\theta}}
  \;=\; \frac{\mathrm{d}{\ell}(\tilde{\mathbf{q}}(\cdot,\bm{\theta}))}{\mathrm{d}\bm{\theta}}\;=\;
  \underbrace{\frac{\partial \ell}{\partial \bm{\theta}}}_{\text{direct dependence} ~=~0}
  + 
  \frac{\partial \ell}{\partial \mathbf{q}}\Bigg\vert_{\mathbf{q}(\cdot) = \tilde{\mathbf{q}}(\cdot, \bm{\theta} )}\cdot
                \frac{\mathrm{d} \tilde{\mathbf{q}}}{\mathrm{d} \bm{\theta}}
               ,
    \label{eq:l_gradient}
\end{equation*}
where $\partial\ell/\partial\bm\theta=0$ because $\ell$ depends on $\bm\theta$ only through the reduced state.
Note that $\partial\ell/\partial\mathbf q$ is in fact a directional derivative (operator) defined on the reduced-state \textit{function space} $\mathcal C^1(\mathcal T;\mathbb R^r)$, rather than on $\mathbb R^r$. 

However, directly evaluating $\mathrm{d}\tilde{\mathbf q}/\mathrm{d}\bm\theta$ is often computationally expensive, because the cost of such primal sensitivity analysis scales with the number of parameters (i.e., $d$ in our setting). Instead we will rewrite the gradient using a continuous adjoint variable that solves a backward ODE, as detailed in Proposition~\ref{thm:adjoint_method} in the next subsection. In particular, the advantage of the adjoint method becomes clear when dealing with a relatively large number of parameters to estimate, as its computational cost is roughly twice the cost of a single primal simulation, irrespective of the number of parameters.

\subsection{The adjoint method}
\label{sec:adjoint_eqs}

We pose the training process to determine the reduced operators $\bm\theta$ as a differential-equation-constrained optimization problem, i.e., 
\begin{equation}
\label{eq:opt_problem}
\begin{aligned}
  \min_{\bm\theta}  \tilde{\ell}(\bm\theta) \quad= \quad& \min_{\bm\theta,\mathbf q(\cdot)}  \ell(\mathbf q(\cdot)) \\
  &\text{s.t.}\quad  \dot{\mathbf q}(t)=\mathbf f\big(\mathbf q(t);\bm\theta\big),\qquad
  \mathbf q(0)=\mathbf q_0,\qquad t\in[0,T],
\end{aligned}
\end{equation}
where $\tilde{\ell}$ and $\ell$ are respectively defined in \eqref{eq:continuous_reduced_loss} and \eqref{eq:continuous_loss}, and
$\mathbf f(\cdot;\bm\theta)$ is the right-hand side of the ROM in \eqref{eq:redopinf}.
Note that the trajectory loss $\ell$ depends only on the reduce-state function $\mathbf q(\cdot)$, measuring its misfit to
the reference trajectory $\mathbf q_{\mathrm{true}}(\cdot)$, and does not depend explicitly on $\bm\theta$.
For this setting, the classical continuous-time adjoint method (see, e.g.,
\cite{bradley2024pde,luchini2024introduction})
yields the following result.
\begin{proposition}[The adjoint method]
    \label{thm:adjoint_method}
    Assume $\mathbf f(\mathbf q;\bm\theta)$ is continuously differentiable in both $\mathbf q$ and $\bm\theta$ and that the initial condition $\mathbf q_0$ is independent
    of $\bm\theta$. Let $\tilde{\mathbf q}(\cdot;\bm\theta)$ solve the state equation in
    \eqref{eq:opt_problem}.
    The gradient of the reduced loss function  \eqref{eq:continuous_reduced_loss} is
    \begin{equation}
        \frac{\mathrm{d}\tilde{\ell}(\bm{\theta})}{\mathrm{d}\bm{\theta}}
   = \int_0^T \bm{\lambda}(t)^{\top}\dfrac{\partial\mathbf{f}(\mathbf{q};\bm{\theta})}{\partial\bm{\theta}}\Bigg\vert_{\mathbf{q}(\cdot) = \tilde{\mathbf{q}}(\cdot; \bm{\theta} ) }~\mathrm{d} t,
        \label{eq:gradient_lagrange}
    \end{equation}
    where the adjoint state variable $\bm\lambda:[0,T]\to\mathbb R^r$ satisfies the
    backward-in-time ODE
    \begin{equation}
        \dot{\bm{\lambda}}(t) = -\left[ \left(\dfrac{\partial \mathbf{f}}{\partial\mathbf{q}}\right)^{\top}\bm{\lambda}(t) + \left( \dfrac{\partial g}{\partial \mathbf{q}} \right)^{\top} \right]\Bigg\vert_{\mathbf{q}(\cdot) = \tilde{\mathbf{q}}(\cdot, \bm{\theta} ) },\quad\bm{\lambda}(T)=\bm{0}.
        \label{eq:adjoint_eqs}
    \end{equation}
    
\end{proposition}
The proof is provided in \ref{app:B}. To evaluate the adjoint equation
\eqref{eq:adjoint_eqs}, one first integrates the forward reduced model to obtain
$\tilde{\mathbf{q}}(t;\bm{\theta})$ on $[0,T]$, and the adjoint ODE \eqref{eq:adjoint_eqs} is then
solved backward in time from $t=T$ to $t=0$. Thereafter, the gradient \eqref{eq:gradient_lagrange} is
computed by numerical integration along the resulting trajectories.

\paragraph{Derivative notation}
We adopt a unified Jacobian advection and use
$\partial_{\mathbf q}$ and $\partial /\partial \mathbf q$ interchangeably.
For a scalar $g:\mathbb R^{r}\!\to\!\mathbb R$,
$\partial_{\mathbf q} g \equiv \nabla_{\!\mathbf q} g^{\top} \in \mathbb R^{1\times r}$.
For a vector field $\mathbf f:\mathbb R^{r}\!\to\!\mathbb R^{r}$,
$\partial_{\mathbf q}\mathbf f \in \mathbb R^{r\times r}$ denotes the Jacobian matrix with
$(\partial_{\mathbf q}\mathbf f)_{ij}=\partial f_i/\partial q_j$.
Similarly, parameter derivatives are $\partial_{\bm\theta}\mathbf f\in\mathbb R^{r\times d}$.

\paragraph{Quadratic ROM: explicit derivatives}
For the quadratic ROM in \eqref{eq:redopinf},
\[
\mathbf f(\mathbf q;\bm\theta)=\mathbf c+\mathbf A\,\mathbf q+\mathbf H(\mathbf q\otimes\mathbf q)+\mathbf B\,\mathbf s(t)
\quad \text{with} \quad
\bm\theta=[\mathbf c,\mathbf A,\mathbf H,\mathbf B],
\]
The derivatives entering
\eqref{eq:gradient_lagrange} and \eqref{eq:adjoint_eqs} are
\begin{equation*}
\begin{aligned}
\partial_{\mathbf q}\mathbf f(\mathbf q;\bm\theta)
&= \mathbf A + 2\mathbf H\big(\mathbf I_r\otimes \mathbf q\big)
\;\in\;\mathbb R^{r\times r},
\\[3pt]
\partial_{\mathbf q} g\big(\mathbf q(t),t\big)
&= 2\big(\mathbf q(t)-\mathbf q_{\mathrm{true}}(t)\big)^\top
\;\in\;\mathbb R^{1\times r},
\\[3pt]
\partial_{\bm\theta}\mathbf f(\mathbf q;\bm\theta)
&=\big[\mathbf I_r,\;
          \mathbf q^{\!\top}\!\otimes \mathbf I_r,\;
          (\mathbf q\otimes \mathbf q)^{\!\top}\!\otimes \mathbf I_r,\;
          \mathbf s(t)^{\!\top}\!\otimes \mathbf I_r\big]
\;\in\;\mathbb R^{r\times d}. 
\end{aligned}
\end{equation*}
Complete derivations are provided in \ref{app:C}.

\subsection{Computational implementation}
Initialization of the parameters $\bm\theta$ is crucial for the nonconvex optimization. We warm-start with the OpInf estimate from the regression in Eq.~\eqref{eq:lst_opinf}. To stabilize this estimator and improve noise robustness, we regularize OpInf with Tikhonov (ridge) regularization and optionally apply truncated SVD (TSVD) to discard noise-dominated directions, exploiting that small singular values largely capture noise whereas large ones carry signal. Hyperparameters (ridge weight and TSVD rank) are chosen by grid search: we fit OpInf on the training set for each setting and select the model that minimizes the validation relative state error (RSE),
\begin{equation}\label{eq:loss_rse}
\mathrm{RSE} \;=\; 
\frac{\big\|\mathbf{q}_{\mathrm{true}}(t)-\mathbf{q}_{\mathrm{pred}}(t)\big\|_2}
     {\big\|\mathbf{q}_{\mathrm{true}}(t)\big\|_2},
\end{equation}
aggregated over the full validation horizon. The selected OpInf estimator provides the initial guess $\bm\theta^{0}$ for adjoint-based training.

In POD, the “energy” of a mode equals the variance it captures; specifically, the squared singular value $\sigma_i^2$ quantifies the variance of mode $i$ \cite{annurev060042}. With i.i.d. Gaussian measurement noise, the noise covariance is isotropic, so the projected noise variance is (approximately) the same in every orthonormal direction, whereas the signal variance concentrates in the leading modes; lower-energy modes thus have lower signal-to-noise ratio (SNR) rather than more absolute noise \cite{101111Tipping}. 
To emphasize informative directions while remaining statistically principled, we use per-mode weights $\omega_i = \sigma_i^{p}/ (\nu_i^2+\tau)$, $\tau=10^{-8}$,
where $\nu_i^2$ is the noise variance projected onto mode $i$. Setting $p=0$ recovers generalized least-squares (inverse-variance) weighting \cite{Strutz}; $p=2$ yields SNR weighting $\sigma_i^2/\nu_i^2$. In this study we adopt $p=1$ as a tempered-SNR choice that prioritizes high-energy modes without over-penalizing the remainder. We estimate $\nu_i^2$ by smoothing each POD coefficient with an adaptive-window Savitzky–Golay filter and taking the residual variance (state minus smoothed state) as $\nu_i^2$ \cite{Savitzky1964, lejarza2022data}. Let $\bm W=\operatorname{diag}(\omega_1,\dots,\omega_r)/\sum_{i=1}^{r}\omega_i$, so that $\operatorname{tr}(\bm W)=1$, which fixes the overall scale and stabilizes gradients across datasets. The weighted pointwise loss is
\[
g(\mathbf q,t)=\big\|\sqrt{\bm W}\big(\mathbf q-\mathbf q_{\mathrm{true}}(t)\big)\big\|_2^2,\]
and since $\bm W$ is diagonal with strictly positive entries, its gradient is
\[
\partial_{\mathbf q} g(\mathbf q,t)=2\,\big(\bm W(\mathbf q(t)-\mathbf q_{\mathrm{true}}(t))\big)^\top.
\]

For large-scale problems one may use stochastic variants such as stochastic gradient descent (SGD) \cite{Ruder2016AnOO}. However, for the smaller datasets considered here, we optimize $\tilde{\ell}(\bm\theta)$ by deterministic gradient descent with Armijo backtracking. The parameter gradient $\nabla_{\bm\theta}\tilde{\ell}$ is computed via the adjoint formula \eqref{eq:gradient_lagrange}, and the update is
\[
\bm\theta^{\,j+1} \;=\; \bm\theta^{\,j} - \eta_j\,\nabla_{\bm\theta}\tilde{\ell}(\bm\theta^{\,j}),
\]
with $\eta_j$ selected by the Armijo condition to ensure sufficient decrease \cite{nocedal1999numerical}. The line search and the full training loop are summarized in Algorithms~\ref{algorithm_1}–\ref{algorithm_2} in \ref{app:A}.

\section{Numerical experiments}

We evaluate the adjoint-based training on three canonical PDEs: the 1D viscous Burgers’ equation, the 2D Fisher–KPP reaction–diffusion equation, and the 2D advection–diffusion equation (ADE). The numerical details for generating the dataset for these examples can be found in \ref{app:D}. These benchmarks stress different aspects of reduced models: Burgers’ equation with small viscosity produces sharp gradients and nonlinear advection–diffusion coupling; Fisher–KPP exhibits front propagation in two spatial dimensions; and the ADE isolates linear transport and diffusion while allowing advection-dominated (high Péclet) regimes.

For each PDE, we generate full-order (FOM) trajectories on a uniform spatial grid up to a fixed horizon $T$, collect state snapshots, and split them chronologically into training/validation/test sets. A POD basis $\mathbf V_r$ is built from the training snapshots and used to project the FOM data to reduced coordinates. Unless stated otherwise, we choose $r$ so that the cumulative snapshot energy satisfies $\kappa_r \ge 0.995$ (i.e., at least $99.5\%$ of the energy retained).

\paragraph{Perturbation protocols}
To assess robustness, we consider two controlled perturbations:
\begin{itemize}
\item \textbf{Snapshot sparsification:} uniformly thinning the time grid (reduced sampling density).
\item \textbf{Additive Gaussian noise:} corrupting the clean snapshots with i.i.d.\ noise
\[
\epsilon(t) ~\sim~ \mathcal N\!\big(0,\;(\delta\,\sigma_{\mathbf q})^2\,\mathbf I\big), \qquad
\delta \in \{0, 40, 80, 120, 160, 200\}\%,
\]
where $\sigma_{\mathbf q}$ is the empirical standard deviation of the clean reduced state over the training window. We refer to $\delta$ as the noise level (NL).
\end{itemize}

\paragraph{Baseline and initialization (OpInf)}
We compare the adjoint-trained ROM against a standard OpInf baseline on the same POD basis and preprocessing. OpInf is implemented with the Python package \texttt{opinf} \cite{opinf_python}. For OpInf regression, we test finite-difference stencils of order 2 (OpInf-ord2) and 6 (OpInf-ord6) to approximate $\dot{\mathbf q}$. We use ridge and TSVD regularization, selecting hyperparameters on the validation set via a grid search over (i) ridge weight $\{0, 10^{-2}, 10^{-1}, 1\}$ and (ii) discarding 1–7 smallest singular directions. The stencil order and regularization are chosen by minimizing the validation RSE (Eq.~\eqref{eq:loss_rse}). The best OpInf estimate initializes the adjoint training, $\bm\theta^{0}$.

\paragraph{Training regime (adjoint)}
When the ODE is integrated end-to-end over a long time horizon, small discretization and modeling errors accumulate and can be amplified by unstable modes of the dynamics, ultimately leading to large prediction errors and ill-conditioned, sometimes exploding gradients. We therefore adopt a multiple-shooting-inspired training scheme with short-horizon rollouts to mitigate exploding gradients in long-time ODE training \cite{Houska2012}.
The trajectory is partitioned into three segments. For each segment, we reset the initial condition to the observed state at the segment start and integrate the ROM only over that short horizon. Parameters are shared across segments, but updates are computed from a single-segment objective (no global sum across segments). We cycle segments with at most 30 adjoint iterations per segment.
During adjoint optimization, we include an ridge penalty on the parameter vector $\bm\theta$. The ridge weight is chosen via a grid search over $\{0, 10^{-2}, 10^{-1}, 1, 10\}$, with model selection by validation-set RSE. 

\paragraph{Optimization and numerics}
Armijo backtracking parameters are fixed to $\alpha=10^{-4}$, $\beta=0.5$, $\gamma=0.5$, and $\eta_0=10^{-3}$; gradient descent is terminated when $\|\nabla_{\bm\theta}\tilde{\ell}\|_2 \le 10^{-8}$. Forward and adjoint ODEs, as well as time integrals, are computed with \texttt{SciPy} routines \cite{virtanen2020scipy}; the data misfit and its gradient are evaluated on the solver’s internal time grid using the temporal interpolant of the observed trajectory.

\subsection{Viscous burgers' equation}
We study the 1D viscous Burgers’ equation
\[
u_t + u\,u_x = \nu\,u_{xx},\qquad x\in[0,1],~t\in[0,T],\quad \nu=0.01,
\]
with homogeneous Dirichlet conditions \(u(t,0)=u(t,1)=0\) and initial condition
\(u(x,0)=\sin(2\pi x)\).
This nonlinear advection–diffusion problem develops steep gradients that are smoothed by viscosity, making it a canonical test for nonlinear model reduction.

\begin{figure}[h]
  \centering
    \centering
    \includegraphics[width=0.5\textwidth]{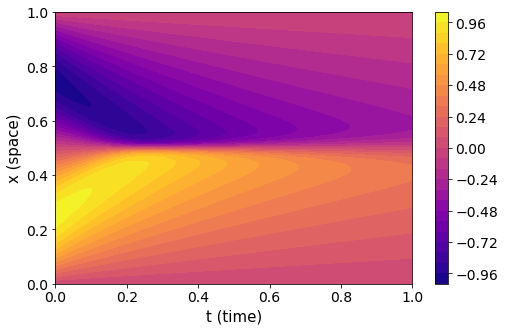}
    \caption{Numerical solution of the viscous Burgers’ equation. Space–time contour of \(u(x,t_i)\).}
  \label{fig:burgers-data}
\end{figure}

We solve the PDE on a uniform spatial grid with \(N=998\) interior points \((\Delta x=1/(N+1))\) and a uniform time grid with \(T=1\), \(M=9999\) steps \((\Delta t=T/M\approx 10^{-4})\). Time stepping uses a second-order semi-implicit (IMEX) finite-difference scheme: Lax–Wendroff for the advective term and Crank–Nicolson for diffusion, enforcing the Dirichlet data at each step. Stacking the discrete solution over all times yields the snapshot matrix \(\mathbf U\in\mathbb{R}^{1000\times 10{,}000}\), which we use for POD and ROM training (see Figure~\ref{fig:burgers-data}).

\begin{figure}[htb]
\centering
\includegraphics[width=\linewidth]{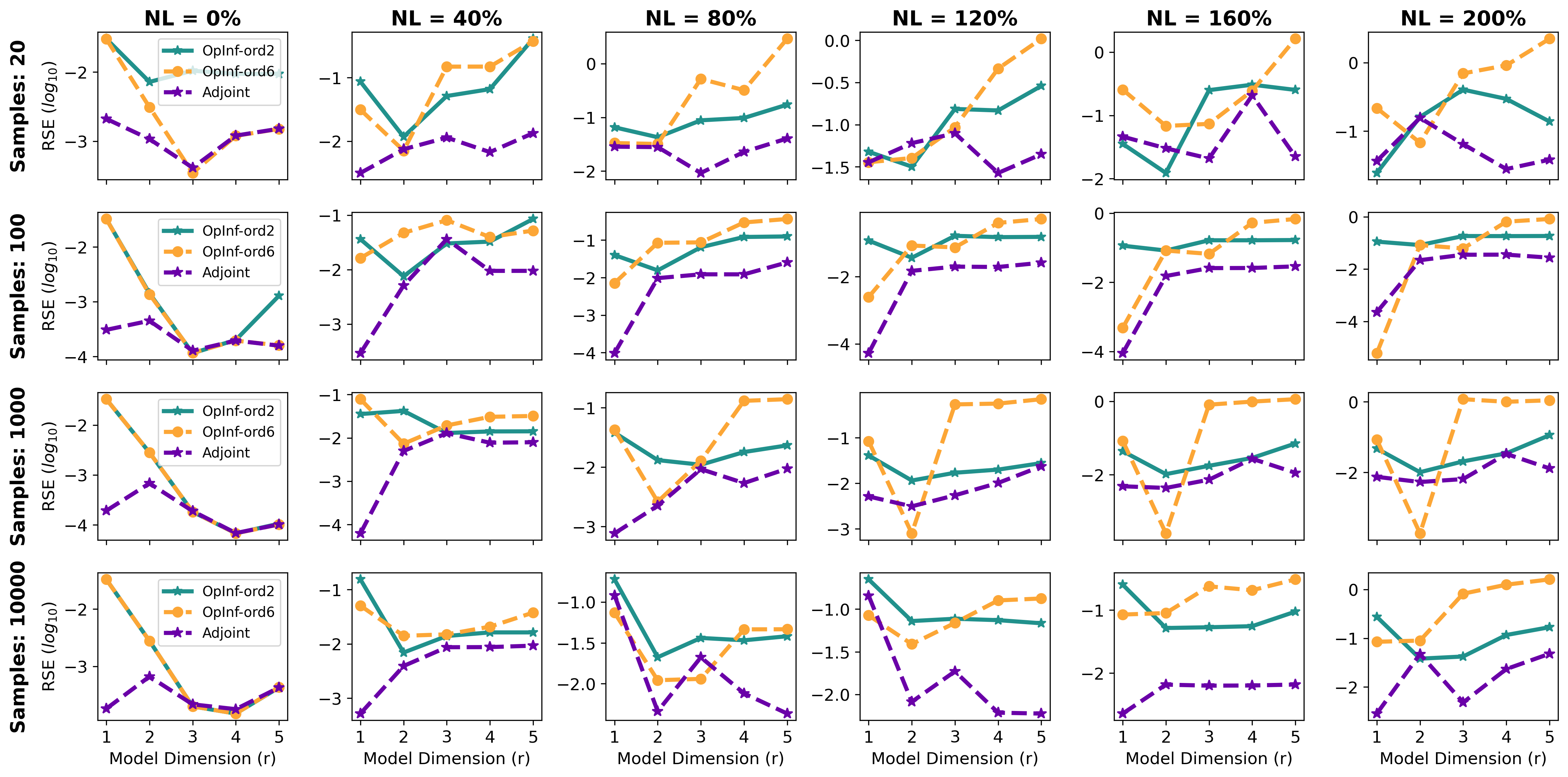}
\caption{Model performance on Burgers’ equation under varying noise and sampling. Columns vary the noise level (NL = 0–200\% of the state standard deviation); rows vary the number of snapshots across train, validation, and test (20, 100, 1000, 10000). Each panel shows test RSE ($\log_{10}$) versus ROM dimension $r$. Methods: Adjoint (ours), OpInf-ord2, and OpInf-ord6. Lower is better.}
\label{fig:plot_error_burgers}
\end{figure}

To probe sampling effects, we uniformly subsample the time grid to obtain 100\% (10000 points), 10\% (1000 points), 1\% (100 points), and 0.2\% (20 points) snapshots. Then we split the trajectory chronologically into train/validation/test as $t\in[0,0.5]$ (50\%), $t\in[0.5,0.6]$ (10\%), and $t\in[0.6,1]$ (40\%). Training and validation snapshots are corrupted with i.i.d. Gaussian noise at levels $\delta\in\{0,40,80,120,160,200\}\%$ of the state standard deviation; test data are clean. For each scenario and ROM dimension $r\in\{1,...,5\}$ (with $\kappa_5\geq0.999$), we (i) fit OpInf-ord2 and OpInf-ord6, selecting ridge/TSVD hyperparameters by validation RSE, and (ii) initialize the adjoint method with the best OpInf estimate. All methods are then rolled out from the test initial condition, and performance is reported as $\log_{10}$-RSE on the test window (Figure~\ref{fig:plot_error_burgers}).

Across sampling and noise conditions, the adjoint-trained ROM attains the lowest test error in most panels (Figure~\ref{fig:plot_error_burgers}). OpInf-ord6 is competitive on clean data (leftmost column) but degrades rapidly as noise increases, which consistent with higher-order differencing amplifying noise. OpInf-ord2 is generally more stable than OpInf-ord6 under noise but still trails the adjoint approach. Notably, at the highest noise level (rightmost column) the adjoint method remains accurate even with only 20 snapshots, while both OpInf variants suffer substantial error. As $r$ increases, errors typically drop until $r\approx 3-4$ and then plateau or worsen in noisy regimes, indicating diminishing returns and potential overfitting.

\begin{figure}[htb]
\centering
\includegraphics[width=.9\linewidth]{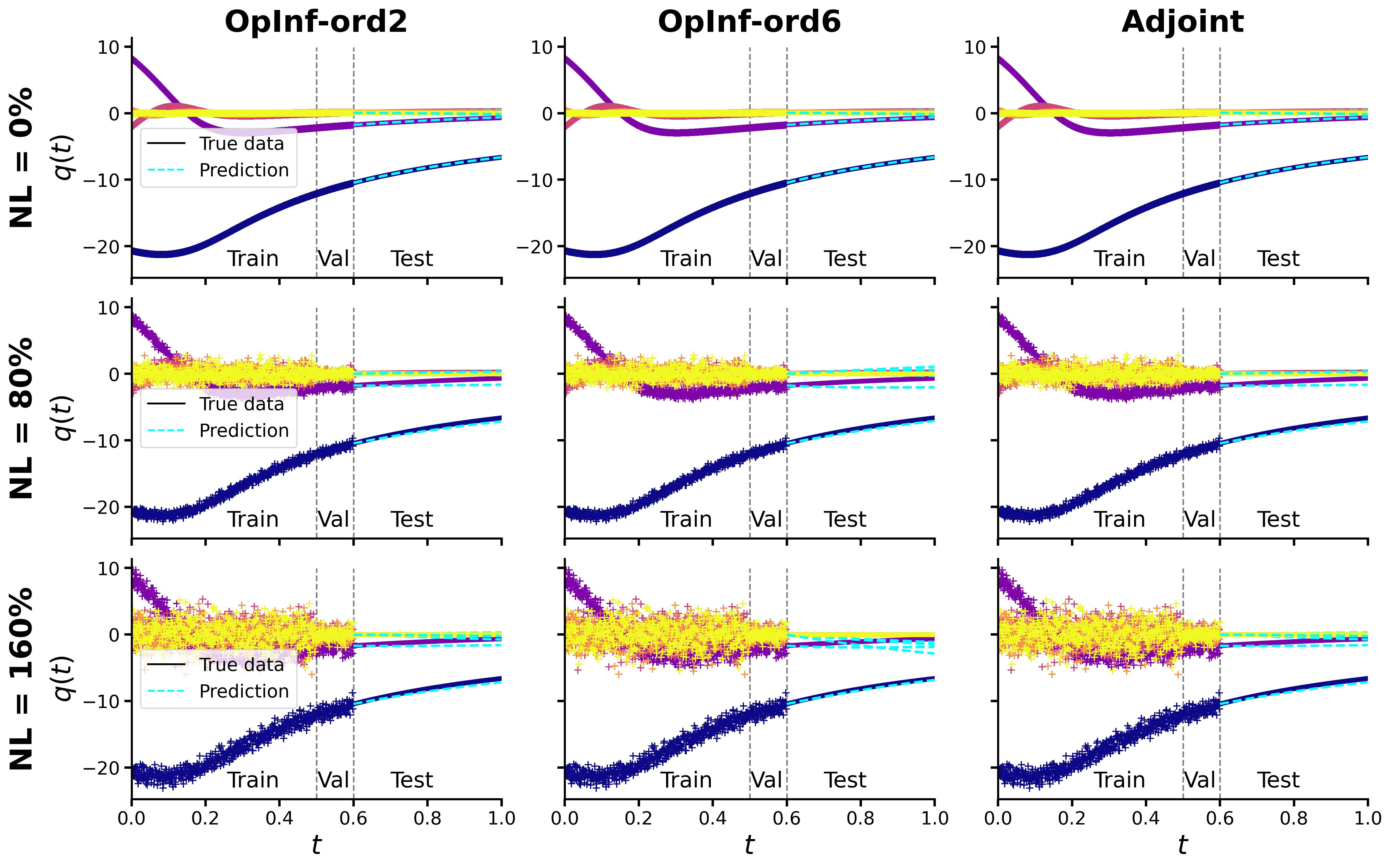}
\caption{Reduced-coordinate rollouts for viscous Burgers’ equation on the test window. ROM dimension $r=5$; $1000$ total snapshots across train, validation, and test. Rows vary the noise level (NL = 0\%, 80\%, 160\%); columns compare OpInf-ord2, OpInf-ord6, and the adjoint method (ours). Gray vertical dashed lines mark train/validation/test splits. Colored markers are the noisy observations $\mathbf q_{\text{true}}(t)$ used for training/validation; solid curves are the clean reference trajectory; cyan dashed curves are model predictions initialized at the test initial condition. Under heavy noise, the adjoint rollout follows the clean trajectory more closely, especially beyond the training window.}
\label{fig:plot_examples_burgers_sample1000}
\end{figure}

\begin{figure}[htb]
\centering
\includegraphics[width=.9\linewidth]{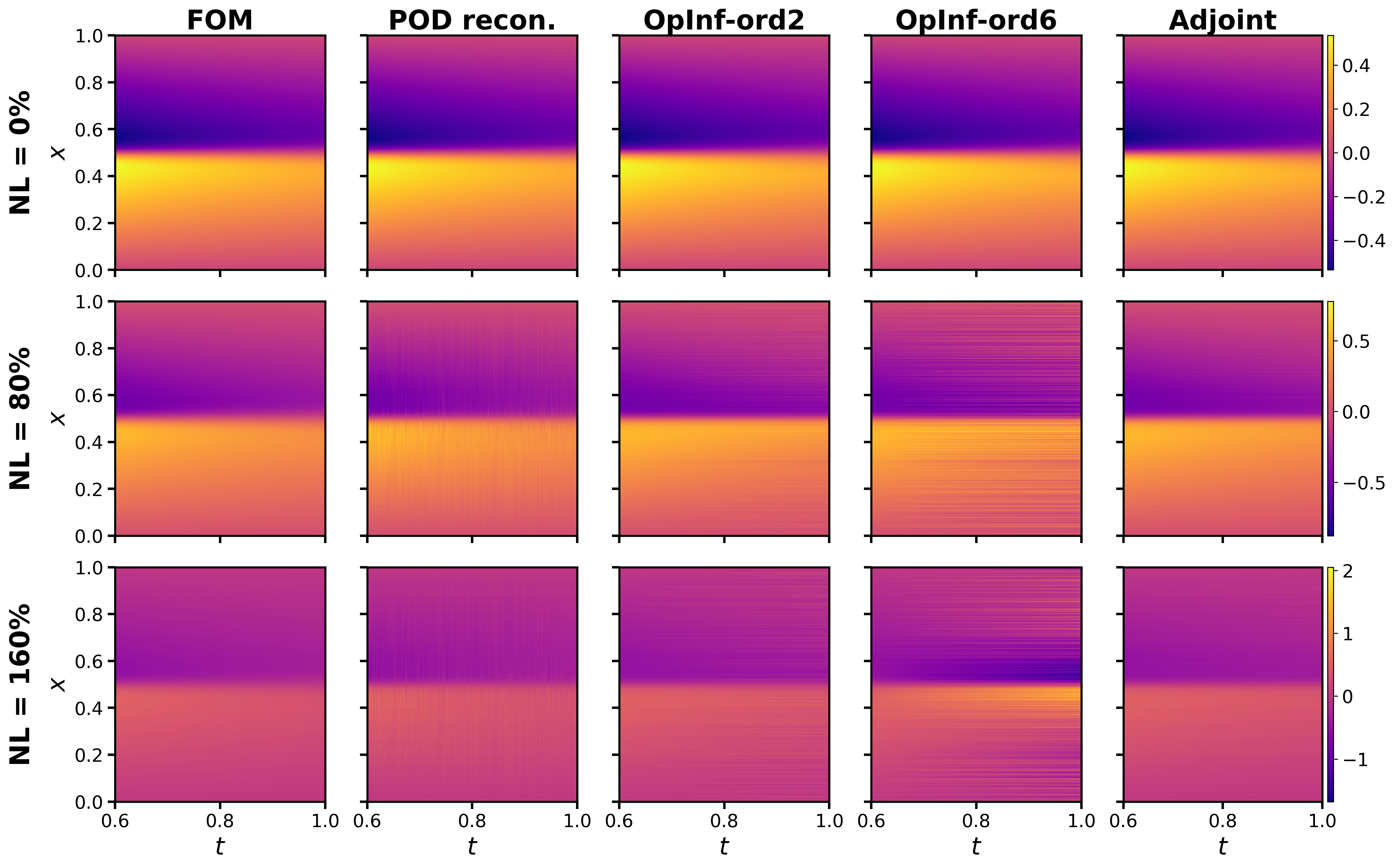}
\caption{
Spatiotemporal evolution $u(x,t)$ for viscous Burgers’ equation on the test window. ROM dimension $r=5$; $1000$ total snapshots across train, validation, and test. Rows vary NL = 0\%, 80\%, 160\%; columns show the FOM and the corresponding reconstruction using POD, OpInf-ord2, OpInf-ord6, and the adjoint ROM. Color maps share a panel-wise scale. The adjoint model best preserves the location and amplitude of the evolving profile under noise, while OpInf-ord6 exhibits pronounced noise imprinting and OpInf-ord2 shows bias and smoothing.}
\label{fig:plot_examples_evolution_burgers_sample1000}
\end{figure}

Figure~\ref{fig:plot_examples_burgers_sample1000} illustrates reduced-coordinate rollouts ($r=5$, 1000 snapshots) at $\delta=\{0,80,160\}\%$: as noise grows, the adjoint ROM consistently tracks the clean trajectory more closely than OpInf,  highlighting its superior gradient fidelity in the presence of data corruption.
Figure~\ref{fig:plot_examples_evolution_burgers_sample1000} shows the corresponding spatiotemporal fields mapped back to the FOM grid; the adjoint model preserves the structure and phase of the evolving profile under heavy noise better than the OpInf baselines.

\subsection{Fisher-KPP equation}

\begin{figure}[htb]
  \centering
    \centering
    \includegraphics[width=\textwidth]{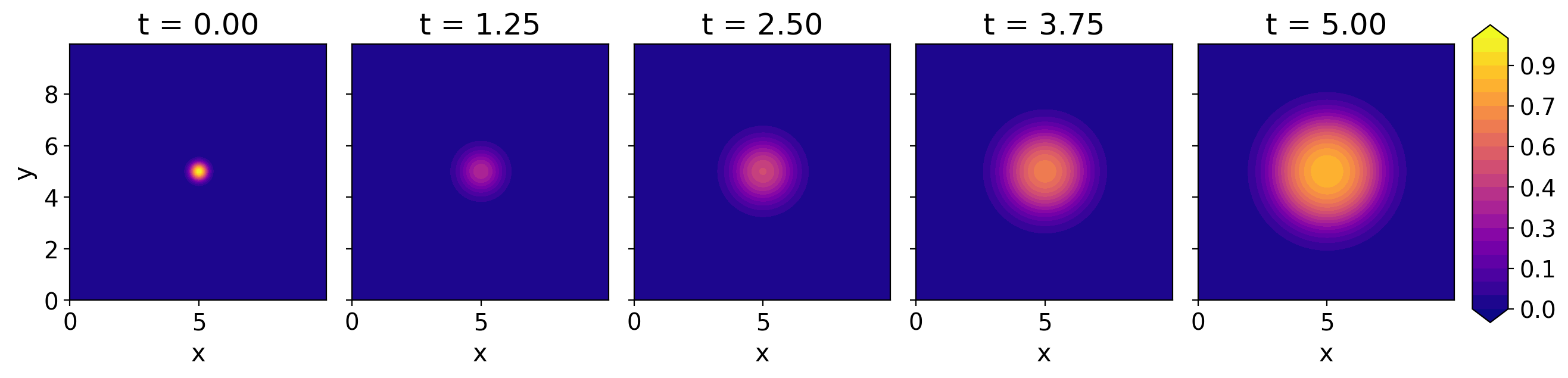}
    \caption{Numerical solution of Fisher-KPP equation. Space–time contour of \(u(x,t_i)\).}
  \label{fig:fkpp-data}
\end{figure}

For this second synthetic experiment, we consider the 2D Fisher–KPP equation
\[
u_t = \mathfrak D\,\big(u_{xx}+u_{yy}\big) + \rho\,u\,(1-u),
\qquad (x,y)\in[0,L_x]\times[0,L_y],~ t\in[0,T],
\]
with homogeneous Neumann (zero–flux) boundary conditions
\(\partial u/\partial n|_{\partial\Omega}=0\) and a Gaussian initial condition centered in the domain,
\[
u(x,y,0)=\exp\!\big(-10\big[(x-L_x/2)^2+(y-L_y/2)^2\big]\big).
\]
We use \(L_x=L_y=10\), \(\mathfrak D=0.1\), and \(\rho=1.0\).
The PDE is solved on a uniform grid with \(N_x=N_y=125\) points in each direction
(\(\Delta x=L_x/(N_x-1)\), \(\Delta y=L_y/(N_y-1)\)) and a uniform time grid with
\(T=5\), \(\Delta t=0.0025\) \((N_t=T/\Delta t=2000)\).
Time stepping is second–order IMEX: Crank–Nicolson for diffusion and explicit Euler for the
logistic reaction term, with Neumann data enforced in the discrete Laplacian.
Stacking the solution over time yields the snapshot matrix
\(\mathbf U\in\mathbb R^{(N_xN_y)\times (N_t+1)}\), used for POD and ROM training
(see Figure~\ref{fig:fkpp-data}).

We follow the same evaluation protocol as in the Burgers' equation case (OpInf initialization, adjoint training, and test-window RSE), with two changes specific to FKPP:
(i) denser sampling baselines via uniform subsampling to 2000 (100\%), 1000 (50\%), 500 (25\%), and 200 (10\%) snapshots;
(ii) a longer horizon and split $t\in[0,3.75]$ (train, 75\%), $t\in[3.75,4.25]$ (validation, 10\%), and $t\in[4.25,5]$ (test, 15\%).

\begin{figure}[htb]
\centering
\includegraphics[width=1\linewidth]{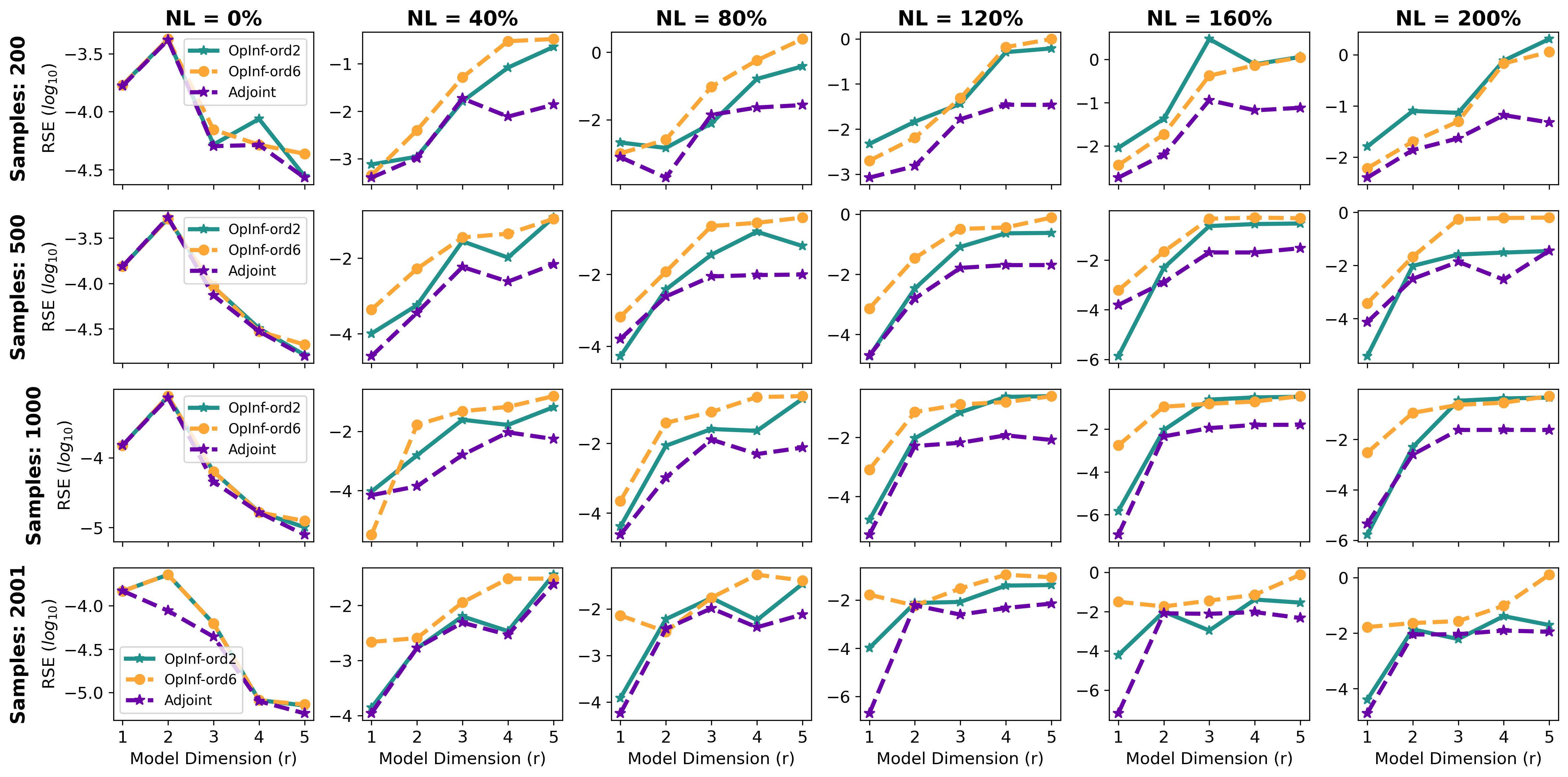}
\caption{Model performance on Fisher-KPP equation under varying noise and sampling. Columns vary the noise level (NL = 0–200\% of the state standard deviation); rows vary the number of snapshots across train, validation, and test (200, 500, 1000, 2001). Each panel shows test RSE ($\log_{10}$) versus ROM dimension $r$. Methods: Adjoint (ours), OpInf-ord2, and OpInf-ord6. Lower is better.}
\label{fig:plot_error_fkpp}
\end{figure}

Figure~\ref{fig:plot_error_fkpp} shows trends broadly consistent with Burgers' equation, with clean data, all methods are comparable and improve with
$r$ up to 5. As the noise level (NL) increases, OpInf-ord6 deteriorates more rapidly than ord2—higher-order differencing amplifies noise, and FKPP’s reaction term preserves those high-frequency artifacts. The adjoint-trained ROM attains the lowest or near-lowest test RSE across most panels—especially with 200–1000 samples and NL~$\geq 80\%$, and still degrades gracefully at NL~$=80\%$. Under noise, gains from increasing $r$ saturate earlier (typically $r=3$ or $4$); OpInf often plateaus or worsens, whereas the adjoint remains stable.

\begin{figure}[htb]
\centering
\includegraphics[width=.9\linewidth]{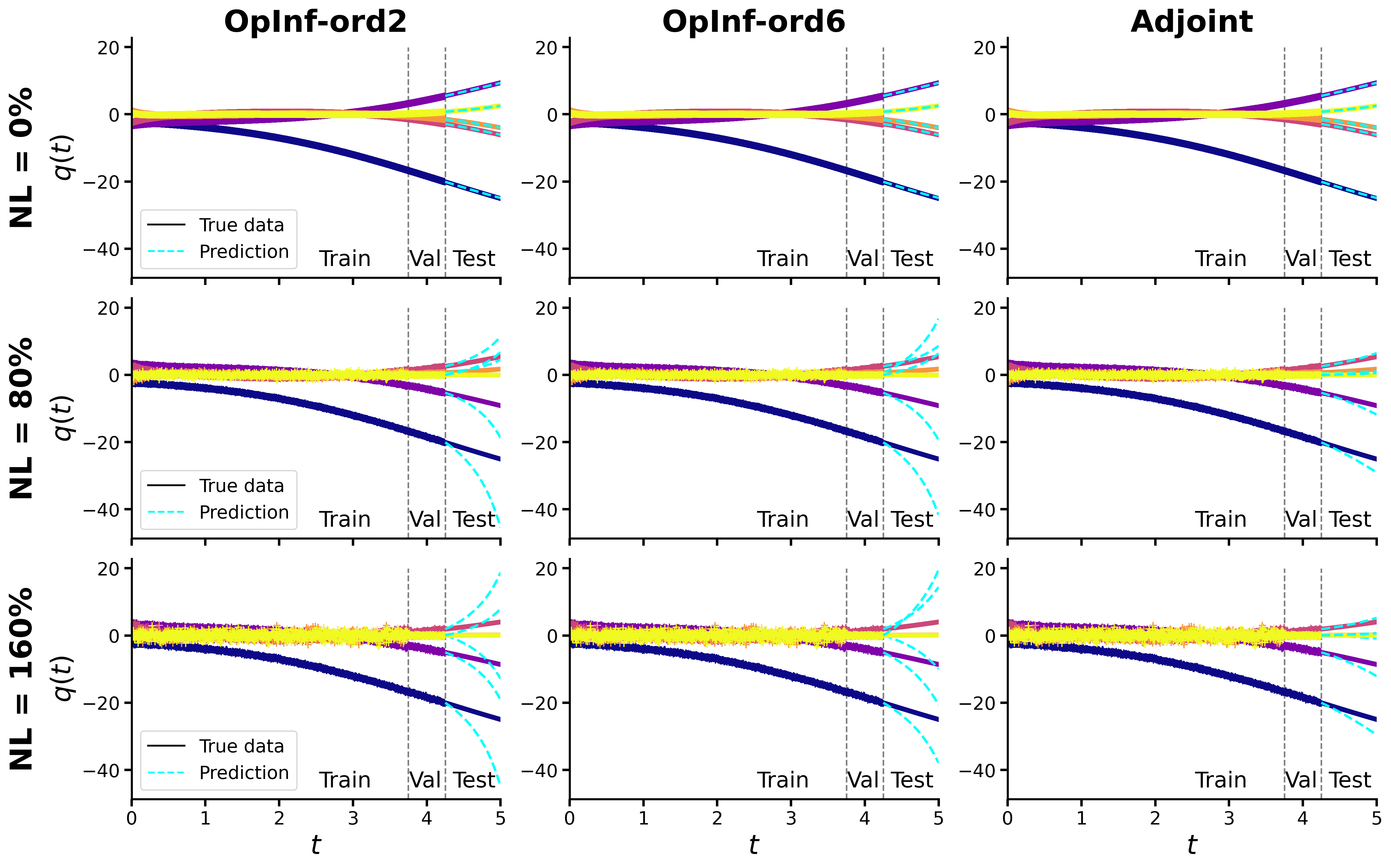}
\caption{Reduced-coordinate rollouts for Fisher-KPP equation on the test window.
ROM dimension $r=5$; $1000$ total snapshots across train, validation, and test. Rows vary the noise level (NL = 0\%, 80\%, 160\%); columns compare OpInf-ord2, OpInf-ord6, and the adjoint method (ours). Gray vertical dashed lines mark train/validation/test splits. Colored markers are the noisy observations $\mathbf q_{\text{true}}(t)$ used for training/validation; solid curves are the clean reference trajectory; cyan dashed curves are model predictions initialized at the test initial condition. Under heavy noise, the adjoint rollout follows the clean trajectory more closely, especially beyond the training window.}
\label{fig:plot_examples_fkpp_sample1000}
\end{figure}

\begin{figure}[htb]
\centering
\includegraphics[width=.9\linewidth]{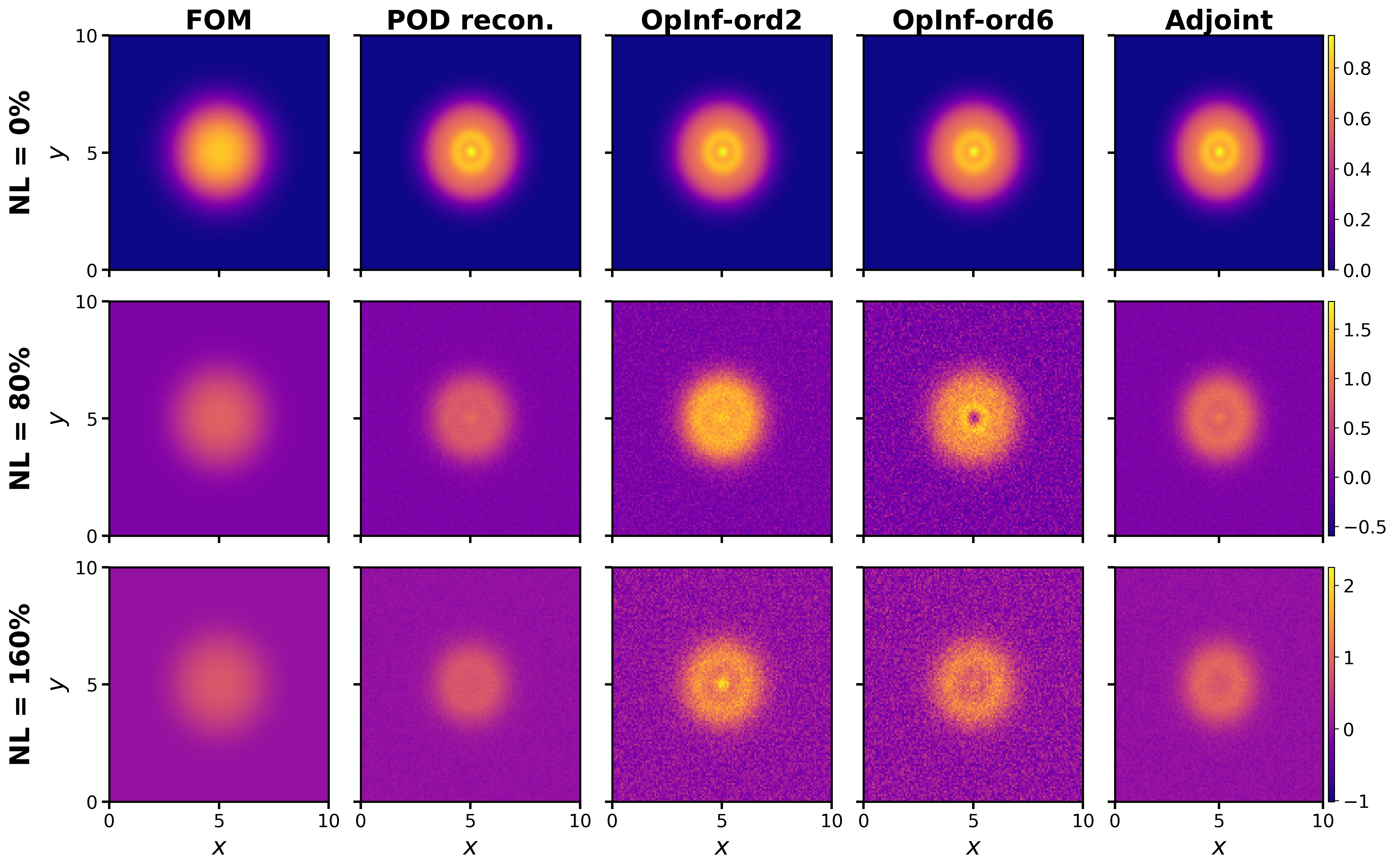}
\caption{Spatial field for Fisher-KPP at $t=0.5$. ROM dimension $r=5$; $1000$ total snapshots across train, validation, and test. Rows vary NL = 0\%, 80\%, 160\%; columns show the FOM and the corresponding reconstruction using POD, OpInf-ord2, OpInf-ord6, and the adjoint ROM. Color maps share a panel-wise scale. The adjoint model best preserves the location and amplitude of the evolving profile under noise, while OpInf-ord6 exhibits pronounced noise imprinting and OpInf-ord2 shows bias and smoothing.}
\label{fig:plot_examples_evolution_fkpp_sample1000}
\end{figure}

In reduced coordinates (Figure~\ref{fig:plot_examples_fkpp_sample1000}), adjoint rollouts stay close to the clean reference throughout the test window; mapped back to the FOM grid at $t=5$ (Figure~\ref{fig:plot_examples_evolution_fkpp_sample1000}), the adjoint preserves the smooth, radially symmetric profile while OpInf exhibits grainy ring artifacts, most pronounced for ord6 at NL~$\geq80\%$.

\subsection{Advection–diffusion equation (ADE)}

We consider the linear advection–diffusion equation
\[
u_t \;=\; c_x\,u_x \;+\; c_y\,u_y \;+\; \nu\,(u_{xx}+u_{yy}),
\qquad (x,y)\in[-1,1]^2,\; t\in[0,T],
\]
with constant advection speeds \(c_x=1\), \(c_y=1.5\) and viscosity \(\nu=0.005\).
We impose periodic boundary conditions in both spatial directions and initialize with a Gaussian,
\[
u(x,y,0) \;=\; \exp\!\Big(-\tfrac{(x-x_0)^2+(y-y_0)^2}{2\sigma^2}\Big),
\quad (x_0,y_0)=(-0.5,-0.5),\;\sigma=0.1.
\]

\begin{figure}[htb]
  \centering
    \centering
    \includegraphics[width=1\textwidth]{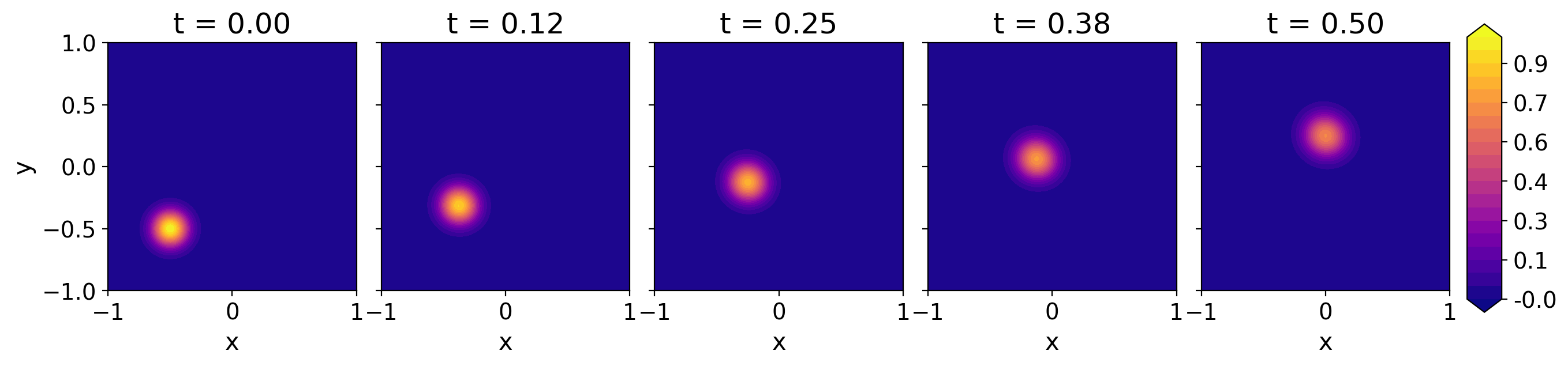}
    \caption{Numerical solution of ADE. Space–time contour of \(u(x,t_i)\).}
  \label{fig:lcd-data}
\end{figure}

The PDE is solved on a uniform grid with \(N_x=N_y=201\) points in each direction
(\(\Delta x=\Delta y=2/(N_x-1)\)) and a uniform time grid with \(T=0.5\), \(\Delta t=10^{-3}\).
Spatial derivatives use second-order centered differences with periodic wrapping; time stepping is explicit
forward Euler. The chosen \(\Delta t\) satisfies standard advective and diffusive Courant–Friedrichs–Lewy (CFL) constraints.
Stacking the solution over time yields the snapshot matrix
\(\mathbf U\in\mathbb R^{(N_xN_y)\times (N_t+1)}\), which we use for POD and ROM training (see Figure \ref{fig:lcd-data}).

We adopt the same evaluation protocol as in the Burgers' equation and FKPP studies (OpInf initialization, adjoint training, test-window RSE), with a shorter horizon $T=0.5$ the split $t\in[0,0.375]$ (train, 75\%), $t\in[0.375,0.425]$ (validation, 10\%), and $t\in[0.425,0.5]$ (test, 15\%), and a broader ROM range $r\in\{1,...,15\}$ (with $\kappa_{15}\geq99.5$) at sample levels $\{200,500,1000,2001\}$.

\begin{figure}[htb]
\centering
\includegraphics[width=1\linewidth]{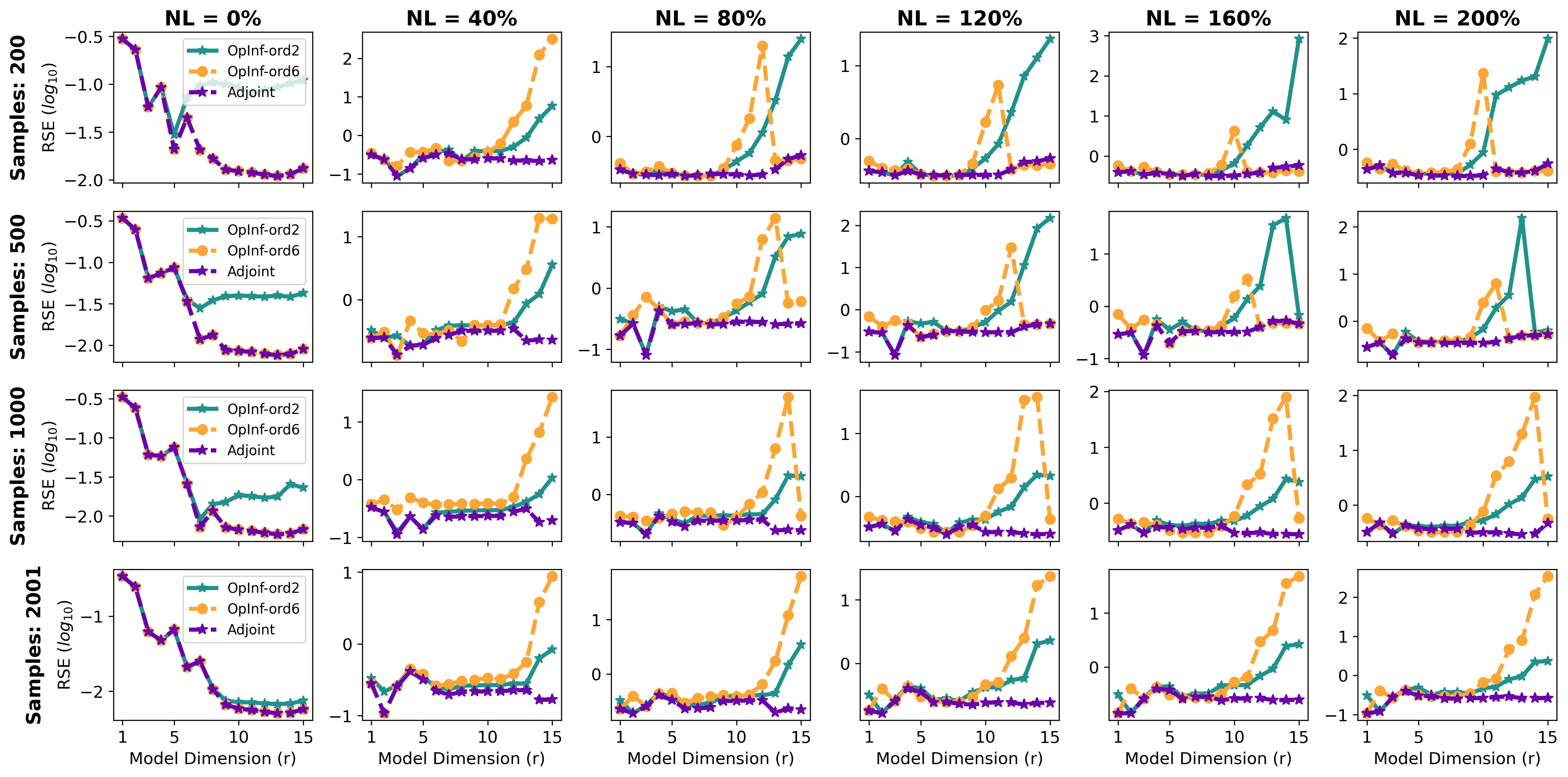}
\caption{Model performance on ADE under varying noise and sampling. Columns vary the noise level (NL = 0–200\% of the state standard deviation); rows vary the number of snapshots across train, validation, and test (200, 500, 1000, 2001). Each panel shows test RSE ($\log_{10}$) versus ROM dimension $r$. Methods: Adjoint (ours), OpInf-ord2, and OpInf-ord6. Lower is better.}
\label{fig:plot_error_ADE}
\end{figure}

Figure~\ref{fig:plot_error_ADE} shows that, with clean data, adjoint training and OpInf-ord6 yield comparable accuracy that improves with $r$. As the noise level increases (NL~$\geq80$), the OpInf baselines become progressively unstable with larger $r$, with pronounced error growth, particularly for ord6 and at lower sample counts, whereas the adjoint method remains stable across dimensions and achieves the lowest or close to the lowest test RSE in most settings. This behavior is consistent with the sensitivity of derivative-based regression to advective noise amplification, which the adjoint formulation avoids.

\begin{figure}[htb]
\centering
\includegraphics[width=.9\linewidth]{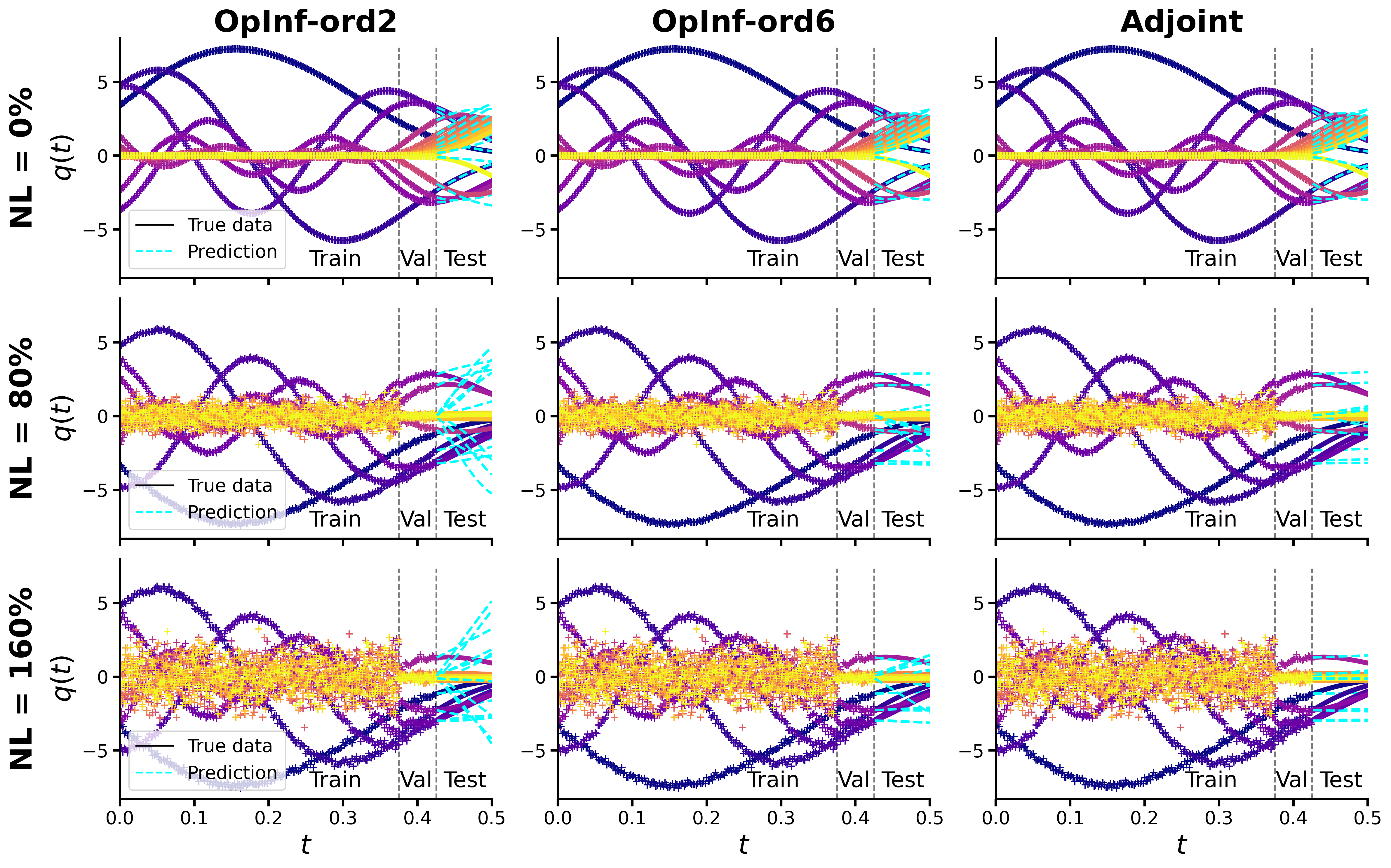}
\caption{Reduced-coordinate rollouts for ADE on the test window.
ROM dimension $r=15$; $1000$ total snapshots across train, validation, and test. Rows vary the noise level (NL = 0\%, 80\%, 160\%); columns compare OpInf-ord2, OpInf-ord6, and the adjoint method (ours). Gray vertical dashed lines mark train/validation/test splits. Colored markers are the noisy observations $\mathbf q_{\text{true}}(t)$ used for training/validation; solid curves are the clean reference trajectory; cyan dashed curves are model predictions initialized at the test initial condition. Under heavy noise, the adjoint rollout follows the clean trajectory more closely, especially beyond the training window.}
\label{fig:plot_examples_ADE_sample1000}
\end{figure}

\begin{figure}[htb]
\centering
\includegraphics[width=.9\linewidth]{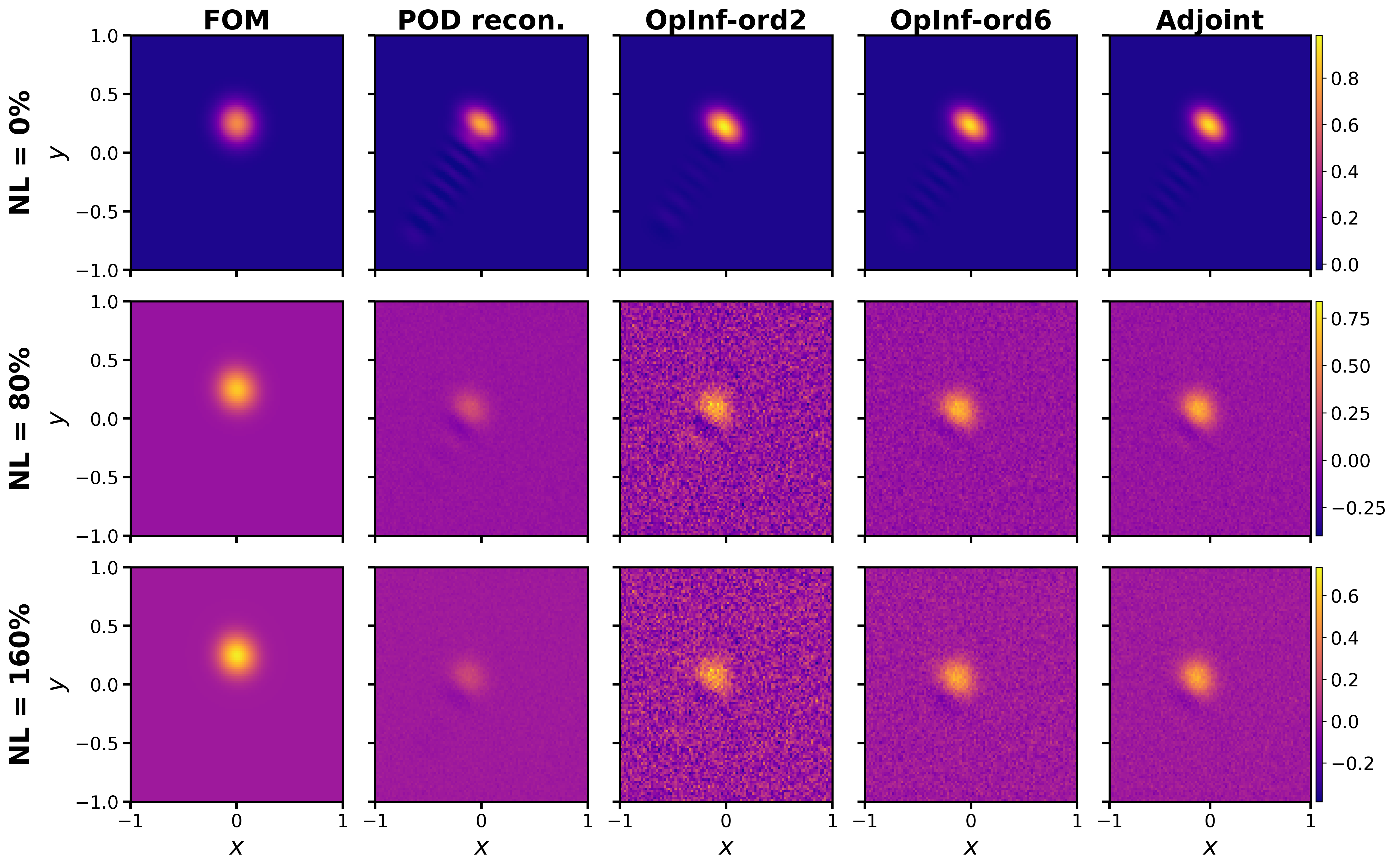}
\caption{Spatial field for ADE at $t=0.5$. ROM dimension $r=15$; $1000$ total snapshots across train, validation, and test. Rows vary NL = 0\%, 80\%, 160\%; columns show the FOM and the corresponding reconstruction using POD, OpInf-ord2, OpInf-ord6, and the adjoint ROM. Color maps share a panel-wise scale. The adjoint model best preserves the location and amplitude of the evolving profile under noise, while OpInf-ord6 exhibits pronounced noise imprinting and OpInf-ord2 shows bias and smoothing.}
\label{fig:plot_examples_evolution_ADE_sample1000_endTrue}
\end{figure}

In reduced coordinates ($r = 15$, $1000$ snapshots; Figure~\ref{fig:plot_examples_ADE_sample1000}), adjoint rollouts remain coherent across the test interval as noise increases, whereas OpInf trajectories drift and fan out after the boundary between training and validation. Reconstructions on the FOM grid at the final time (Figure~\ref{fig:plot_examples_evolution_ADE_sample1000_endTrue}) show that the adjoint model preserves the smooth, transported Gaussian profile, while OpInf predictions exhibit speckled, advected artifacts, particularly for OpInf-ord6. Overall, the ADE results are consistent with the Burgers' equation and FKPP cases, with a sharper separation between methods in high noise, advection-dominated regimes.

\section{Concluding remarks}
We introduced an adjoint-based training framework for quadratic ROMs that minimizes a continuous-time trajectory loss and computes exact gradients via a backward adjoint solve, thereby avoiding the error-prone time-derivative estimates required by standard operator inference, which becomes particularly unreliable under sparse temporal sampling and measurement noise. In the proposed adjoint-based training, each optimization step costs one forward and one backward time integration, independent of the number of model parameters.  

Across three representative PDE benchmarks, the adjoint-trained ROM consistently matched or improved upon standard OpInf on clean data and delivered markedly better robustness under both perturbation protocols: reduced snapshot density via uniform subsampling and additive Gaussian noise at increasing levels. In particular, the adjoint formulation remained stable in regimes where high-order finite-difference OpInf (ord6) degraded rapidly, and it typically outperformed the more noise-tolerant low-order baseline (ord2) as noise increased or data became sparse. These results support the central premise that fitting trajectories in continuous time, rather than matching estimated derivatives, yields more reliable reduced operators when training data are imperfect. The main limitations are reliance on a polynomial model class, sensitivity to the accuracy of temporal discretization, and greater wall time than least squares regression. 

Several extensions are natural directions for future work. For example, incorporating structure-preserving constraints into the training scheme could further improve long-horizon generalization and stability. 
Moreover, extending the proposed method to handle parametric variation in physical properties and multiple trajectories by varied initial conditions, as well as assessing robustness under non-Gaussian and correlated noise, would improve applicability to experimental settings and sensor data integration. Overall, the proposed adjoint-based strategy provides a practical and flexible way for learning polynomial ROMs that are substantially less sensitive to sparse and noisy data.

\section*{Declaration of competing interest}
The authors declare that they have no conflict of interest.


\section*{Data availability}
Codes to
reproduce the experiments are available at \url{https://github.com/lindliu/adjoint_opinf}.

\clearpage
\appendix
\section{Algorithms}
\label{app:A}

\begin{algorithm}[H]
\caption{Armijo backtracking line search (steepest descent)}\label{algorithm_1}
\SetAlgoLined
\KwIn{current iterate $\bm\theta^{j}$; objective $\tilde\ell$; parameters $\alpha\in(0,1)$, $\beta\in(0,1)$, $\gamma\in(0,1)$; initial step $\eta_{0}>0$; max backtracks $N_{\max}$}
\KwOut{next iterate $\bm\theta^{j+1}$, updated seed $\eta_{0}$, accepted step $\eta_{j}$}
\BlankLine
Set $\mathscr{G}^{j} \gets \nabla \tilde\ell(\bm\theta^{j})$;\quad $\eta \gets \eta_{0}$;\quad $\text{accepted} \gets \text{false}$\;
\For{$i=1,\dots,N_{\max}$}{
  \If{$\tilde\ell(\bm\theta^{j} - \eta\,\mathscr{G}^{j}) \le \tilde\ell(\bm\theta^{j}) - \alpha\,\eta\,\|\mathscr{G}^{j}\|_2^{2}$}{
     $\text{accepted}\gets \text{true}$\;
     $\bm\theta^{j+1}\gets \bm\theta^{j} - \eta\,\mathscr{G}^{j}$\;
     $\eta_{j}\gets \eta$\;
     \Return
  }
  $\eta \gets \beta\,\eta$ \tcp*{backtrack}
}
\If{not $\text{accepted}$}{
  $\bm\theta^{j+1}\gets \bm\theta^{j}$;\quad $\eta_{0}\gets \gamma\,\eta_{0}$;\quad $\eta_{j}\gets 0$\;
  \Return
}
\end{algorithm}

\begin{algorithm}[H]
\caption{Adjoint-based training of ROM parameters}\label{algorithm_2}
\SetAlgoNoLine

\KwIn{FOM snapshots $\mathbf U$; time window $[t_1,t_k]$; Armijo params $(\alpha,\beta,\gamma,\eta_0)$; tolerance $\epsilon$; max iters $J_{\max}$; POD rank $r$; weight exponent $p$}
\KwOut{optimized parameters $\bm\theta^\ast$}

\BlankLine
\textbf{POD projection:}
Compute POD basis $\mathbf V_r$ (energy threshold or fixed $r$) and reduced snapshots
$[\mathbf q_{\mathrm{true}}(t_1),\dots,\mathbf q_{\mathrm{true}}(t_k)] \gets \mathbf V_r^\top \mathbf U$.
Let $\{\lambda_i\}_{i=1}^r$ be the squared singular values\;

\textbf{Per-state weights:}
Estimate noise variances $\nu_i^2$ via Savitzky--Golay smoothing of each state and residual variance.
Set $\omega_i \gets \sigma_i^{\,p}/\nu_i^2$ (default $p=1$ is a tempered-SNR) and
$\bm W \gets \operatorname{diag}(\omega_1,\dots,\omega_r)$\;

\textbf{Initialization:}
Obtain $\bm\theta^{0}$ from OpInf (Eq.~\eqref{eq:lst_opinf}) with ridge/TSVD hyperparameters selected on a validation set\;

\For{$j=0,1,\dots,J_{\max}-1$}{
  \textbf{Forward solve:}
  Integrate the reduced ODE \eqref{eq:redopinf} on $[t_1,t_k]$ with parameters $\bm\theta^{j}$ to obtain
  $\tilde{\mathbf q}(t;\bm\theta^{j})$\;

  \textbf{Weighted residual:}
  For each $t$, set $\partial_{\mathbf q} g(t)\gets 2\,W\big(\tilde{\mathbf q}(t;\bm\theta^{j})-\mathbf q_{\mathrm{true}}(t)\big)$\;

  \textbf{Adjoint solve:}
  Integrate \eqref{eq:adjoint_eqs} backward from $t_k$ to $t_1$ with terminal condition $\bm\lambda(t_k)=\mathbf 0$
  using $\partial_{\mathbf q}\, g(t)$ and $\partial_{\mathbf q}\,\mathbf f$. Interpolate $\partial_{\mathbf q}\, g(t)$ on the solver grid\;

  \textbf{Parameter gradient:}
  Compute $\nabla_{\bm\theta}\tilde\ell(\bm\theta^{j})$ via \eqref{eq:gradient_lagrange}\;

  \textbf{Line search \& update:}
  Apply Armijo backtracking (Alg.~\ref{algorithm_1}) to obtain $\bm\theta^{j+1}$ and the updated seed $\eta_0$\;

  \textbf{Stopping:} \If{$\|\nabla_{\bm\theta}\tilde\ell(\bm\theta^{j+1})\|_2 \le \epsilon$}{\text{break}}
}
\Return $\bm\theta^\ast \gets \bm\theta^{j+1}$.
\end{algorithm}

\section{Proof of Proposition~\ref{thm:adjoint_method}}
\label{app:B}

Proposition~\ref{thm:adjoint_method} follows standard adjoint arguments (see, e.g., \cite{Antil2018,bradley2024pde}). We adapt them to our setting.

\subsection{Proof by Lagrangian formulation}
\begin{proof}

The associated Lagrangian to the constrained optimization \eqref{eq:opt_problem} is
\begin{equation*}
  \mathcal L\big(\mathbf q(\cdot),\bm\theta,\bm\lambda(\cdot)\big)
  := \ell\big(\mathbf q(\cdot)\big)
     + \int_0^T \bm\lambda(t)^\top\!\left[\mathbf f(\mathbf q(t);\bm\theta)-\dot{\mathbf q}(t)\right]\,dt
     - \bm\lambda(0)^\top\!\big(\mathbf q(0)-\mathbf q_0\big).
\end{equation*}
Because the initial condition $\mathbf q(0)=\mathbf q_0$ is independent of $\bm\theta$, one can simplify the Lagrangian as
\[
\mathcal L({\mathbf q}(\cdot;\bm\theta),\bm\theta,\bm\lambda(\cdot))
:= \int_0^T \Big( g({\mathbf q},t) \;+\; \bm\lambda(t)^\top\!\big(\mathbf f({\mathbf q};\bm\theta)-\dot{{\mathbf q}}\big) \Big)\,\mathrm dt.
\]
Since the ODE constraint holds for $\tilde{\mathbf q}(\cdot;\bm\theta)$, then for any
$\bm\lambda(\cdot)\in\mathcal C^1(\mathcal T;\mathbb R^r)$,
\begin{equation*}
  \mathcal L\big(\tilde{\mathbf q}(\cdot;\bm\theta),\bm\theta,\bm\lambda(\cdot)\big)
  = \tilde{\ell}(\bm\theta).
\end{equation*}

Under the stated smoothness, differentiation under the integral and integration by parts are valid. Differentiating with respect to $\bm\theta$ while treating $\bm\lambda$ as an auxiliary function gives
\[
\frac{\mathrm d}{\mathrm d\bm{\theta}}\tilde{\ell}(\bm{\theta})=\frac{\mathrm d}{\mathrm d\bm\theta}\mathcal L(\tilde{\mathbf q}(\cdot;\bm\theta),\bm\theta,\bm\lambda(\cdot))
=\int_0^T \!\left[
\underbrace{\frac{\partial g}{\partial\bm\theta}}_{=\,0}
+ \frac{\partial g}{\partial\tilde{\mathbf q}}\,\frac{\partial \tilde{\mathbf q}}{\partial\bm\theta}
+ \bm\lambda(t)^\top\!\left(\frac{\partial \mathbf f}{\partial\bm\theta}
+ \frac{\partial \mathbf f}{\partial \tilde{\mathbf q}}\,\frac{\partial \tilde{\mathbf q}}{\partial\bm\theta}
- \frac{\mathrm d}{\mathrm dt}\frac{\partial \tilde{\mathbf q}}{\partial\bm\theta}\right)
\right]\mathrm dt.
\]
Integrating the last term by parts gives
\[
-\!\int_0^T \bm\lambda^\top \frac{\mathrm d}{\mathrm dt}\!\Big(\frac{\partial \tilde{\mathbf q}}{\partial\bm\theta}\Big)\,\mathrm dt
= -\,\bm\lambda^\top \frac{\partial \tilde{\mathbf q}}{\partial\bm\theta}\Big|_{0}^{T}
+ \int_0^T \dot{\bm\lambda}^\top \frac{\partial \tilde{\mathbf q}}{\partial\bm\theta}\,\mathrm dt.
\]
Since $\tilde{\mathbf q}(0)$ does not depend on $\bm\theta$, $\frac{\partial \tilde{\mathbf q}}{\partial\bm\theta}(0)=\mathbf 0$, and thus
\begin{equation}
\frac{\mathrm d}{\mathrm d\bm\theta}\tilde{\ell}(\bm{\theta})
= \int_0^T \!\Big(
\bm\lambda^\top \frac{\partial \mathbf f}{\partial \bm\theta}
+ \underbrace{\big[\tfrac{\partial g}{\partial \tilde{\mathbf q}}
+ \bm\lambda^\top\tfrac{\partial \mathbf f}{\partial \tilde{\mathbf q}}
+ \dot{\bm\lambda}^\top \big]\frac{\partial \tilde{\mathbf q}}{\partial \bm\theta}}_{{\text{sensitivity term}}}
\Big)\,\mathrm dt
\;-\; \bm\lambda(T)^\top \frac{\partial \tilde{\mathbf q}}{\partial \bm\theta}(T).
\label{eq:ppp}
\end{equation}
Here, one can choose the adjoint $\bm\lambda$ to eliminate the sensitivity term by imposing
\[
\dot{\bm\lambda}(t)
= -\Big(\frac{\partial \mathbf f}{\partial\tilde{\mathbf q}}\Big)^{\!\top}\bm\lambda(t)
  - \Big(\frac{\partial g}{\partial \tilde{\mathbf q}}\Big)^{\!\top},
\qquad
\bm\lambda(T)=\mathbf 0,
\]
which makes both the bracketed term and the terminal boundary term vanish, and hence gives the adjoint ODE system. Eventually, \eqref{eq:ppp} simplifies to
\[
\frac{\mathrm d}{\mathrm d\bm\theta}\tilde{\ell}(\bm{\theta})
= \int_0^T \bm\lambda(t)^\top \frac{\partial \mathbf f(\tilde{\mathbf q}(t);\bm\theta)}{\partial \bm\theta}\,\mathrm dt,
\]
completing the proof.

\end{proof}

\subsection{Proof by primal sensitivity analysis}
We include this alternative proof here as it offers additional insight into the problem through the primal sensitivity analysis.

\begin{proof}
We start with the \emph{primal sensitivity} equations for the state solution $\tilde{\mathbf{q}}(t,\bm{\theta})$ and parameters $\theta_i$, $i=1,\dots,d$:\\
$$\dot{\tilde{\mathbf{q}}}(t)
= \mathbf{f}\bigl(\tilde{\mathbf{q}}(t,\bm{\theta}),\bm{\theta}\bigr),
\qquad
\partial_{\theta_i}\dot{\tilde{\mathbf{q}}}(t)
= \frac{\partial \mathbf{f}}{\partial \tilde{\mathbf{q}}}
  \bigl(\tilde{\mathbf{q}}(t,\bm{\theta}),\bm{\theta}\bigr)
  \,\partial_{\theta_i}\tilde{\mathbf{q}}
+ \partial_{\theta_i}\mathbf{f}
  \bigl(\tilde{\mathbf{q}},\bm{\theta}\bigr).$$
Multiplying the second equation by an arbitrary test function $\mathbf{v}(\cdot)\in\mathcal{C}^1(\mathcal{T};\mathbb{R}^r)$, integrating from $0$ to $T$, and rearranging, we can rewrite the primal problem in its weak form\\
$$\int_{0}^{T}
\Biggl[
  \partial_{\theta_i}\dot{\tilde{\mathbf{q}}}(t,\bm{\theta})
  - \biggl(\dfrac{\partial \mathbf{f}}{\partial \tilde{\mathbf{q}}}\biggr)\,
    \partial_{\theta_i}\tilde{\mathbf{q}}(t,\bm{\theta})
\Biggr]^{\!\top}
\mathbf{v}(t)\,\mathrm{d}t
=
\int_{0}^{T}
\bigl(\partial_{\theta_i}\mathbf{f}\bigr)^{\!\top}
\mathbf{v}(t)\,\mathrm{d}t,$$
or, in operator notation,\\
$$\langle \mathscr{L}_{\bm \theta}~\partial_{\theta_i}\tilde{\mathbf{q}},\,\mathbf{v}\rangle
= \langle \partial_{\theta_i}\mathbf{f},\,\mathbf{v}\rangle,$$
where the operator $\mathscr{L}_{\bm \theta}: \mathcal{C}^1(\mathcal{T};\mathbb R^r)\to\mathcal{C}(\mathcal{T};\mathbb R^r)$ is defined as
$$(\mathscr{L}_{\bm \theta}\mathbf{w})(t) = \dot{\mathbf{w}}(t) - \dfrac{\partial \mathbf{f}}{\partial \tilde{\mathbf{q}}}(t,{\bm \theta})\mathbf{w}(t), \quad\mathbf{w}(\cdot)\in\mathcal{C}^1(\mathcal{T};\mathbb R^r),$$
and $\langle \cdot,\cdot\rangle$ denotes the usual $L_2$-inner product.

On the other hand, the gradient of the loss $\tilde{\ell}$ with respect to $\theta_i$ is\\
$$\frac{\mathrm{d}}{\mathrm{d}\theta_i}\,\tilde\ell(\bm{\theta})
= \int_{0}^{T}
\frac{\partial g}{\partial \tilde{\mathbf{q}}}\bigl(\tilde{\mathbf{q}}(t,\bm{\theta}),t\bigr)
\,\partial_{\theta_i}\tilde{\mathbf{q}}(t,\bm{\theta})\,\mathrm{d}t
= \langle \nabla_{\tilde{\mathbf{q}}}\,g,\,\partial_{\theta_i}\tilde{\mathbf{q}}\rangle.$$
Now, we define the adjoint problem by introducing an adjoint variable $\bm{\lambda}(\cdot)\in\mathcal{C}^1(\mathcal{T};\mathbb R^r)$ such that
$$\langle \mathbf{w},\,{\mathscr{L}^*_{\bm \theta}}\,\bm{\lambda}\rangle
= \langle\mathbf{w}, \nabla_{\tilde{\mathbf{q}}}\,g\,\rangle\,,
\quad\forall\,\mathbf{w}(\cdot)\in\mathcal{C}^1(\mathcal{T},\mathbb R^r)\text{ with }\mathbf{w}(0)=\bm{0},$$
where ${\mathscr{L}^*_{\bm \theta}}$ is the adjoint operator of $\mathscr{L}_{\bm \theta}$.  By the definition of an adjoint operator, this is equivalent to\\
$$\langle \mathscr{L}_{\bm \theta}\,\mathbf{w},\,\bm{\lambda}\rangle
= \langle \nabla_{\tilde{\mathbf{q}}}\,g,\,\mathbf{w}\rangle.$$
Choosing $\mathbf{w} = \partial_{\theta_i}\tilde{\mathbf{q}}$ in this identity immediately gives\\
$$\frac{\mathrm{d}}{\mathrm{d}\theta_i}\,\tilde\ell(\bm \theta)
= \bigl\langle \mathscr{L}_{\bm \theta}\,\partial_{\theta_i}\tilde{\mathbf{q}},\,\bm{\lambda}\bigr\rangle
\underset{
  \substack{
    \text{primal problem}\\
    \text{with }\mathbf{v}\,=\,\bm{\lambda}
  }
}{=}
\bigl\langle \partial_{\theta_i}\mathbf{f},\,\bm{\lambda}\bigr\rangle
= \int_{0}^{T}
  \bm{\lambda}^{\!\top}\,
  \partial_{\theta_i}\mathbf{f}(t,\bm{\theta})\,
  \mathrm{d} t.$$
To derive the adjoint ODE itself, we expand the weak form of the adjoint problem as
$$\int_{0}^{T}
\Biggl[
  \dot{\mathbf{w}}(t)
  - \dfrac{\partial\mathbf{f}}{\partial\tilde{\mathbf{q}}}
    \mathbf{w}(t)
\Biggr]^{\!\top}
\bm{\lambda}(t)\,\mathrm{d}t
= \int_{0}^{T}
\dfrac{\partial g}{\partial\tilde{\mathbf{q}}}
\mathbf{w}(t)\,\mathrm{d}t.$$
Integrating by parts gives that
$$\int_{0}^{T}\dot{\mathbf{w}}(t)^{\!\top}\bm{\lambda}(t)\,\mathrm{d} t =  \left[ \mathbf{w}(t)^{\top}\bm{\lambda}(t)\right]_0^T - \int_0^T \dot{\bm{\lambda}}(t)^{\top}\mathbf{w}(t)~\mathrm{d} t.$$
Using $\mathbf{w}(0)=\bm{0}$ and imposing the adjoint terminal condition $\bm{\lambda}(T)=\bm{0}$ to kill the boundary terms, we obtain\\
$$\int_{0}^{T}
\bigl[
  -\,\dot{\bm{\lambda}}(t)
\bigr]^{\!\top}
\mathbf{w}(t)\,\mathrm{d}t
=
\int_{0}^{T}
\Biggl[
  \biggl(\dfrac{\partial\mathbf{f}}{\partial\tilde{\mathbf{q}}}\biggr)^{\!\top}\bm{\lambda}(t)
  + \biggl(\dfrac{\partial g}{\partial\tilde{\mathbf{q}}}\biggr)^{\!\top}
\Biggr]^{\!\top}
\mathbf{w}(t)\,\mathrm{d}t.$$
Since $\mathbf{w}$ is arbitrary, the integrands on both sides should coincide, which yields the final expression for the adjoint equations\\
$$\dot{\bm{\lambda}}(t)
= -\Biggl[
    \biggl(\dfrac{\partial\mathbf{f}}{\partial\tilde{\mathbf{q}}}\biggr)^{\!\top}\bm{\lambda}(t)
    + \biggl(\dfrac{\partial g}{\partial\tilde{\mathbf{q}}}\biggr)^{\!\top}
  \Biggr],
\quad
\bm{\lambda}(T)=\bm{0}.$$
\end{proof}

\section{Differential identities}
\label{app:C}

We use directional derivatives $\mathcal D h(\mathbf q)[\mathbf v]
:= \lim_{\varepsilon\to0}\frac{h(\mathbf q+\varepsilon\mathbf v)-h(\mathbf q)}{\varepsilon}$.
For vector fields $h:\mathbb R^r\to\mathbb R^r$, the Jacobian $\partial_{\mathbf q}h(\mathbf q)$
is the unique linear map satisfying $\mathcal D h(\mathbf q)[\mathbf v]=\big(\partial_{\mathbf q}h(\mathbf q)\big)\mathbf v$
for all $\mathbf v\in\mathbb R^r$.

\paragraph{Setup}
For the quadratic ROM
\[
\mathbf f(\mathbf q;\bm\theta)=\mathbf c+\mathbf A\,\mathbf q
+\mathbf H(\mathbf q\otimes\mathbf q)+\mathbf B\,\mathbf s(t),
\quad
\bm\theta=[\mathbf c,\mathbf A,\mathbf H,\mathbf B],
\]
we compute the derivatives appearing in \eqref{eq:adjoint_eqs}–\eqref{eq:gradient_lagrange}.

\paragraph{State Jacobian $\partial_{\mathbf q}\mathbf f$}
For any direction $\mathbf v\in\mathbb R^r$,
\[
\mathbf f(\mathbf q+\varepsilon\mathbf v;\bm\theta)
= \mathbf c + \mathbf A\mathbf q + \varepsilon\,\mathbf A\mathbf v
  + \mathbf H(\mathbf q\otimes\mathbf q)
  + \varepsilon\,\mathbf H(\mathbf q\otimes\mathbf v + \mathbf v\otimes\mathbf q)
  + \mathbf B\mathbf s + \mathcal O(\varepsilon^2),
\]
so
\[
\mathcal D\mathbf f(\mathbf q)[\mathbf v]
= \mathbf A\,\mathbf v
  + \mathbf H(\mathbf v\otimes\mathbf q)
  + \mathbf H(\mathbf q\otimes\mathbf v)
= \big(\mathbf A + \mathbf H(\mathbf I_r\!\otimes\!\mathbf q)
               + \mathbf H(\mathbf q\!\otimes\!\mathbf I_r)\big)\mathbf v.
\]
Hence,
\[
\ \partial_{\mathbf q}\mathbf f(\mathbf q;\bm\theta)
= \mathbf A + \mathbf H(\mathbf I_r\!\otimes\!\mathbf q) + \mathbf H(\mathbf q\!\otimes\!\mathbf I_r)
\in\mathbb R^{r\times r}. 
\]
Since the quadratic tensor is symmetric in its last two modes,
$\mathbf H_{i,jk}=\mathbf H_{i,kj}$ (i.e., $\mathbf H(\mathbf q\otimes\mathbf v)=\mathbf H(\mathbf v\otimes\mathbf q)$
for all $\mathbf q,\mathbf v$), this simplifies to
$\partial_{\mathbf q}\mathbf f=\mathbf A+2\,\mathbf H(\mathbf I_r\!\otimes\!\mathbf q)$.

\paragraph{Loss gradient $\partial_{\mathbf q}g$}
For $g(\mathbf q,t)=\|\mathbf q-\mathbf q_{\mathrm{true}}(t)\|_2^2$,
\[
\mathcal Dg(\mathbf q)[\mathbf v]=2\big(\mathbf q-\mathbf q_{\mathrm{true}}(t)\big)^\top\mathbf v
\quad\Rightarrow\quad
\partial_{\mathbf q}g(\mathbf q,t)=2\big(\mathbf q-\mathbf q_{\mathrm{true}}(t)\big)^\top\in\mathbb R^{1\times r}.
\]
With a diagonal weight $\bm W$, $g=\|\bm W^{1/2}(\mathbf q-\mathbf q_{\mathrm{true}})\|_2^2$ yields
$\partial_{\mathbf q}g=2\,\big(\bm W(\mathbf q-\mathbf q_{\mathrm{true}})\big)^\top$.

\paragraph{Parameter derivatives $\partial_{\bm\theta}\mathbf f$}
Let $\mathrm{vec}(\cdot)$ stack matrix columns. Using
$\mathrm{vec}(\mathbf A\mathbf q)=(\mathbf q^\top\!\otimes \mathbf I_r)\,\mathrm{vec}(\mathbf A)$ and
$\mathrm{vec}\!\big(\mathbf H(\mathbf q\otimes\mathbf q)\big)
=((\mathbf q\otimes \mathbf q)^\top\!\otimes \mathbf I_r)\,\mathrm{vec}(\mathbf H)$, we obtain
\[
\partial_{\bm\theta}\mathbf f(\mathbf q;\bm\theta)
=\big[\,\underbrace{\mathbf I_r}_{\partial\mathbf f/\partial\mathbf c}\;\;,\;\;
          \underbrace{\mathbf q^{\!\top}\!\otimes \mathbf I_r}_{\partial\mathbf f/\partial\operatorname{vec}(\mathbf A)}\;\;,\;\;
          \underbrace{(\mathbf q\otimes \mathbf q)^{\!\top}\!\otimes \mathbf I_r}_{\partial\mathbf f/\partial\operatorname{vec}(\mathbf H)}\;\;,\;\;
          \underbrace{\mathbf s(t)^{\!\top}\!\otimes \mathbf I_r}_{\partial\mathbf f/\partial\operatorname{vec}(\mathbf B)}\,\big]
\in\mathbb R^{r\times d},
\]
with $d=r+r^2+r^3+rm$.

\section{Numerical Discretization Details}
\label{app:D}

\subsection{Discretization for Viscous Burgers' Equation}

We discretize on a uniform grid in space and time. Let \(x_j=j\,\Delta x\) for
\(j=0,\dots,N+1\) with \(N=998\) and \(\Delta x=1/(N+1)\). Denote
\(\mathbf u^m=(u_0^m,\dots,u_{N+1}^m)^\top \approx u(\cdot,t^m)\) at times
\(t^m=m\,\Delta t\), \(m=0,\dots,M\), where \(T=1\), \(M=9999\), and \(\Delta t=T/M\approx 10^{-4}\).
Homogeneous Dirichlet values are enforced as \(u_0^m=u_{N+1}^m=0\).

\paragraph{Second-order semi-implicit time stepping} 
Convection is advanced explicitly by a Lax–Wendroff
update to an intermediate field \(\mathbf w^m\) (with \(w_0^m=w_{N+1}^m=0\)). For
\(j=1,\dots,N\),
\[
\begin{aligned}
w_j^m =\, & u_j^m
- \frac{\Delta t}{2\,\Delta x}\,u_j^m\,(u_{j+1}^m - u_{j-1}^m) \\
&+ \frac{\Delta t^2}{2\,\Delta x^2}\,u_j^m
\Big[\tfrac{1}{2}(u_{j+1}^m - u_{j-1}^m)^2
+ u_j^m\,(u_{j+1}^m - 2u_j^m + u_{j-1}^m)\Big].
\end{aligned}
\]
Diffusion is treated implicitly with the trapezoidal (Crank–Nicolson) rule using the discrete
Laplacian \( \mathbf T_{\Delta x}=\Delta x^{-2}\,\mathrm{tridiag}\{1,-2,1\}\) (first/last rows
modified to enforce Dirichlet boundaries). The semi-implicit update reads
\[
\Bigl(\mathbf I - \tfrac{\nu\Delta t}{2}\,\mathbf T_{\Delta x}\Bigr)\mathbf u^{m+1}
= \mathbf w^m + \tfrac{\nu\Delta t}{2}\,\mathbf T_{\Delta x}\,\mathbf u^{m},
\]
which we solve each step with a tridiagonal/direct solver on interior nodes, then restore boundary values~\cite{0078096}.

\paragraph{Snapshots} 
Stacking the time states yields the snapshot matrix
\(\mathbf U=[\mathbf u^0~\cdots~\mathbf u^M]\in\mathbb R^{(N+2)\times(M+1)}\), which we use to build the POD basis and train the ROMs.

\subsection{Discretization for Fisher-KPP Equation}

Let the spatial domain be \(\Omega=[0,L_x]\times[0,L_y]\) with \(L_x=L_y=10\).
We use uniform grids \(x_i=i\,\Delta x\), \(i=0,\dots,N_x-1\), and
\(y_j=j\,\Delta y\), \(j=0,\dots,N_y-1\), where \(N_x=N_y=125\),
\(\Delta x=L_x/(N_x-1)\), \(\Delta y=L_y/(N_y-1)\).
Let \(\mathbf u^m\in\mathbb R^{N_xN_y}\) stack \(u(x_i,y_j,t^m)\) at \(t^m=m\Delta t\),
with \(T=5\), \(\Delta t=0.005\), \(m=0,\dots,N_t\), \(N_t=1000\).

\paragraph{Discrete Laplacian with Neumann BCs}
Construct 1D second–difference matrices \(\mathbf L_x^{\mathrm{1D}}\in\mathbb R^{N_x\times N_x}\) and
\(\mathbf L_y^{\mathrm{1D}}\in\mathbb R^{N_y\times N_y}\) with interior stencil
\(\frac{1}{\Delta x^2}\{1,-2,1\}\) and \(\frac{1}{\Delta y^2}\{1,-2,1\}\), and
zero–flux (Neumann) enforcement at the first/last rows via the standard mirrored–point
modification (e.g.\ diagonal \(-2/\Delta x^2\) with off–diagonal \(+2/\Delta x^2\) at the boundary).
The 2D Laplacian is the Kronecker sum
\[
\mathbf L \;=\; I_{N_y}\otimes \mathbf L_x^{\mathrm{1D}} \;+\; \mathbf L_y^{\mathrm{1D}}\otimes I_{N_x}
\;\in\; \mathbb R^{(N_xN_y)\times(N_xN_y)}.
\]

\paragraph{Time stepping by the IMEX Crank–Nicolson–Euler method}
One step of the semi–implicit scheme reads
\[
\frac{\mathbf u^{m+1}-\mathbf u^{m}}{\Delta t}
= \frac{\mathfrak D}{2}\,\mathbf L(\mathbf u^{m+1}+\mathbf u^{m})
\;+\; \rho\,\mathbf u^{m}\circ(\mathbf 1-\mathbf u^{m}),
\]
i.e.,
\[
(I - \alpha \mathbf L)\,\mathbf u^{m+1}
= (I + \alpha \mathbf L)\,\mathbf u^{m} \;+\; \Delta t\,\rho\,\mathbf u^{m}\circ(\mathbf 1-\mathbf u^{m}), 
\]
where \(\circ\) denotes the Hadamard product and \(\alpha:=\tfrac{\mathfrak D\,\Delta t}{2}\). At each time level, we solve the sparse linear
system for \(\mathbf u^{m+1}\) (e.g.\ with a sparse direct solver).

\paragraph{Snapshots}
The snapshot matrix is \(\mathbf U=[\mathbf u^0~\cdots~\mathbf u^{N_t}]\in\mathbb R^{(N_xN_y)\times (N_t+1)}\)
and is used to build the POD basis and train the ROMs.

\subsection{Discretization for ADE}

Let the domain be \(\Omega=[-1,1]\times[-1,1]\) with periodic boundaries.
We use uniform grids \(x_i=-1+i\,\Delta x\), \(i=0,\dots,N_x-1\), and
\(y_j=-1+j\,\Delta y\), \(j=0,\dots,N_y-1\), with \(N_x=N_y=201\),
\(\Delta x=\Delta y=2/(N_x-1)\).
Denote by \(\mathbf u^m\in\mathbb R^{N_xN_y}\) the lexicographically stacked state
at \(t^m=m\Delta t\) with \(T=0.5\), \(\Delta t=10^{-3}\), \(m=0,\dots,N_t\) \((N_t=500)\).

\paragraph{Periodic spatial differences}
For \(u_{i,j}^m \approx u(x_i,y_j,t^m)\), centered differences with periodic wrap
(indices modulo \(N_x,N_y\)) are
\[
\begin{aligned}
(D_x u)^m_{i,j} &= \frac{u^m_{i+1,j}-u^m_{i-1,j}}{2\Delta x}, \qquad
(D_y u)^m_{i,j} = \frac{u^m_{i,j+1}-u^m_{i,j-1}}{2\Delta y},\\
(\Delta_h u)^m_{i,j} &= \frac{u^m_{i+1,j}-2u^m_{i,j}+u^m_{i-1,j}}{\Delta x^2}
                     + \frac{u^m_{i,j+1}-2u^m_{i,j}+u^m_{i,j-1}}{\Delta y^2}.
\end{aligned}
\]

\paragraph{Time stepping by the explicit Euler method}
One time step is
\[
u^{m+1}_{i,j}
= u^{m}_{i,j}
+ \Delta t\Big[-\,c_x (D_x u)^m_{i,j}
               -\,c_y (D_y u)^m_{i,j}
               + \nu\,(\Delta_h u)^m_{i,j}\Big],
\]
applied for all \(i,j\) with periodic index wrapping.

\paragraph{CFL condition for stability}
A sufficient condition for stabilizing the explicit scheme is
\[
\frac{|c_x|\Delta t}{\Delta x}+\frac{|c_y|\Delta t}{\Delta y}\le 1,
\qquad
\Delta t\,\nu\!\left(\frac{1}{\Delta x^2}+\frac{1}{\Delta y^2}\right)\le \frac{1}{2},
\]
which is satisfied by the parameters set above.

\paragraph{Snapshots}
\(\mathbf U=[\mathbf u^0~\cdots~\mathbf u^{N_t}]\in\mathbb R^{(N_xN_y)\times (N_t+1)}\) is used for POD and ROM training.




\bibliographystyle{elsarticle-num}
\bibliography{references}

@Book{quarteroni2015reduced,
  author    = {Quarteroni, Alfio and Manzoni, Andrea and Negri, Federico},
  publisher = {Springer},
  title     = {{R}educed {B}asis {M}ethods for {P}artial {D}ifferential {E}quations: {A}n {I}ntroduction},
  year      = {2016},
  address   = {Cham, Switzerland},
}

@Article{ghattas2021learning,
  author    = {Ghattas, Omar and Willcox, Karen E.},
  journal   = {Acta Numerica},
  title     = {{L}earning physics-based models from data: {P}erspectives from inverse problems and model reduction},
  year      = {2021},
  pages     = {445--554},
  volume    = {30},
  doi       = {https://doi.org/10.1017/S0962492921000064},
  publisher = {Cambridge University Press},
}

@Article{chen2018neural,
  author  = {Chen, Ricky T. Q. and Rubanova, Yulia and Bettencourt, Jesse and Duvenaud, David},
  journal = {Advances in Neural Information Processing Systems},
  title   = {{N}eural ordinary differential equations},
  year    = {2018},
  volume  = {31},
  doi     = {https://doi.org/10.48550/arXiv.1806.07366},
}

@InCollection{Antil2018,
  author    = {Harbir Antil and Dmitriy Leykekhman},
  booktitle = {Frontiers in PDE-Constrained Optimization},
  publisher = {Springer},
  title     = {{A} brief introduction to {PDE}-constrained optimization},
  year      = {2018},
  pages     = {3-40},
  doi       = {https://doi.org/10.1007/978-1-4939-8636-1_1},
}

@Article{bradley2024pde,
  author      = {Andrew M. Bradley},
  journal = {Technical Report, Stanford University},
  title       = {{PDE}-constrained optimization and the adjoint method},
  year        = { 2024}
}

@Article{kramer2024learning,
  author    = {Kramer, Boris and Peherstorfer, Benjamin and Willcox, Karen E.},
  journal   = {Annual Review of Fluid Mechanics},
  title     = {Learning Nonlinear Reduced Models from Data with Operator Inference},
  year      = {2024},
  number    = {1},
  pages     = {521--548},
  volume    = {56},
  doi       = {https://doi.org/10.1146/annurev-fluid-121021-025220},
  publisher = {Annual Reviews},
}

@Article{luchini2024introduction,
  author  = {Luchini, Paolo and Bottaro, Alessandro},
  journal = {arXiv preprint arXiv:2404.17304},
  title   = {{A}n {I}ntroduction to {A}djoint {P}roblems},
  year    = {2024},
  doi     = {https://doi.org/10.48550/arXiv.2404.17304},
}

@Book{nocedal1999numerical,
  author    = {Nocedal, Jorge and Wright, Stephen J.},
  publisher = {Springer},
  title     = {{N}umerical {O}ptimization},
  year      = {1999},
  address   = {New York, NY},
}

@article{Ruder2016AnOO,
author = {Bottou, L\'{e}on and Curtis, Frank E. and Nocedal, Jorge},
title = {Optimization Methods for Large-Scale Machine Learning},
journal = {SIAM Review},
volume = {60},
number = {2},
pages = {223-311},
year = {2018},
doi = {https://doi.org/10.1137/16M1080173
}
}

@Article{berkooz1993proper,
  author    = {Berkooz, Gal and Holmes, Philip and Lumley, John L.},
  journal   = {Annual review of fluid mechanics},
  title     = {The Proper Orthogonal Decomposition in the Analysis of Turbulent Flows},
  year      = {1993},
  number    = {1},
  pages     = {539--575},
  volume    = {25},
  doi       = {https://doi.org/10.1146/annurev.fl.25.010193.002543},
  publisher = {Annual Reviews 4139 El Camino Way, PO Box 10139, Palo Alto, CA 94303-0139, USA},
}

@Article{virtanen2020scipy,
  author  = {Virtanen, Pauli and Gommers, Ralf and Oliphant, Travis E. and
            Haberland, Matt and Reddy, Tyler and Cournapeau, David and
            Burovski, Evgeni and Peterson, Pearu and Weckesser, Warren and
            Bright, Jonathan and {van der Walt}, St{\'e}fan J. and
            Brett, Matthew and Wilson, Joshua and Millman, K. Jarrod and
            Mayorov, Nikolay and Nelson, Andrew R. J. and Jones, Eric and
            Kern, Robert and Larson, Eric and Carey, C J and
            Polat, {\.I}lhan and Feng, Yu and Moore, Eric W. and
            {VanderPlas}, Jake and Laxalde, Denis and Perktold, Josef and
            Cimrman, Robert and Henriksen, Ian and Quintero, E. A. and
            Harris, Charles R. and Archibald, Anne M. and
            Ribeiro, Ant{\^o}nio H. and Pedregosa, Fabian and
            {van Mulbregt}, Paul and {SciPy 1.0 Contributors}},
  title   = {{{SciPy} 1.0: Fundamental algorithms for scientific
            computing in python}},
  journal = {Nature Methods},
  year    = {2020},
  volume  = {17},
  pages   = {261--272},
  adsurl  = {https://rdcu.be/b08Wh},
  doi     = {https://doi.org/10.1038/s41592-019-0686-2},
}

@article{benner2015survey,
author = {Benner, Peter and Gugercin, Serkan and Willcox, Karen},
title={A Survey of Projection-based Model Reduction Methods for Parametric Dynamical Systems},
journal = {SIAM Review},
volume = {57},
number = {4},
pages = {483-531},
year = {2015},
doi = {https://doi.org/10.1137/130932715}
}

@article{lejarza2022data,
  title={Data-driven discovery of the governing equations of dynamical systems via moving horizon optimization},
  author={Lejarza, Fernando and Baldea, Michael},
  journal={Scientific Reports},
  volume={12},
  number={1},
  pages={1--15},
  year={2022},
  publisher={Springer}
}

@Article{Savitzky1964,
author={Savitzky, Abraham.
and Golay, M. J. E.},
title={Smoothing and Differentiation of Data by Simplified Least Squares Procedures.},
journal={Analytical Chemistry},
year={1964},
month={Jul},
day={01},
publisher={American Chemical Society},
volume={36},
number={8},
pages={1627-1639},
issn={0003-2700},
doi={https://doi.org/10.1021/ac60214a047}
}

@article{annurev060042,
   author = "Rowley, Clarence W. and Dawson, Scott T.M.",
   title = "Model Reduction for Flow Analysis and Control", 
   journal= "Annual Review of Fluid Mechanics",
   year = "2017",
   volume = "49",
   number = "Volume 49, 2017",
   pages = "387-417",
   doi = "https://doi.org/10.1146/annurev-fluid-010816-060042",
   publisher = "Annual Reviews",
   issn = "1545-4479"
  }

@article{101111Tipping,
    author = {Tipping, Michael E. and Bishop, Christopher M.},
    title = {Probabilistic Principal Component Analysis},
    journal = {Journal of the Royal Statistical Society Series B: Statistical Methodology},
    volume = {61},
    number = {3},
    pages = {611-622},
    year = {2002},
    month = {01},
    issn = {1369-7412},
    doi = {https://doi.org/10.1111/1467-9868.00196}
}

@book{Strutz,
author = {Strutz, Tilo},
title = {Data Fitting and Uncertainty: A Practical Introduction to Weighted Least Squares and Beyond},
year = {2010},
isbn = {3834810223},
publisher = {Vieweg and Teubner},
address = {Wiesbaden, DEU}
}

@article{doi:10.1137/090771806,
author = {Halko, N. and Martinsson, P. G. and Tropp, J. A.},
title = {Finding Structure with Randomness: Probabilistic Algorithms for Constructing Approximate Matrix Decompositions},
journal = {SIAM Review},
volume = {53},
number = {2},
pages = {217-288},
year = {2011},
doi = {https://doi.org/10.1137/090771806}
}

@Article{Hansen1987,
author={Hansen, Per Christian},
title={The truncatedSVD as a method for regularization},
journal={BIT Numerical Mathematics},
year={1987},
month={Dec},
day={01},
volume={27},
number={4},
pages={534-553},
issn={1572-9125},
doi={https://doi.org/10.1007/BF01937276},
}

@article{PEHERSTORFER2016196,
title = {Data-driven operator inference for nonintrusive projection-based model reduction},
journal = {Computer Methods in Applied Mechanics and Engineering},
volume = {306},
pages = {196-215},
year = {2016},
issn = {0045-7825},
doi = {https://doi.org/10.1016/j.cma.2016.03.025},
author = {Benjamin Peherstorfer and Karen Willcox}}

@misc{opinf_python,
  author       = {Shane A. McQuarrie and {Willcox Research Group}},
  title        = {opinf: Operator Inference in Python},
  year         = {2025},
  howpublished = {\url{https://pypi.org/project/opinf/}},
  note         = {Python package version 0.5.16}
}

@article{f2890069d4,
 ISSN = {00401706},
 author = {Arthur E. Hoerl and Robert W. Kennard},
 journal = {Technometrics},
 number = {1},
 pages = {80--86},
 publisher = {[Taylor \& Francis, Ltd., American Statistical Association, American Society for Quality]},
 title = {Ridge Regression: Biased Estimation for Nonorthogonal Problems},
 urldate = {2025-12-04},
 volume = {42},
 year = {2000}
}

@article{Willoughby,
author = {Willoughby, Ralph A.},
title = {Solutions of ill-posed problems ({A}. {N}. {T}ikhonov and {V}. {Y}. {A}rsenin)},
journal = {SIAM Review},
volume = {21},
number = {2},
pages = {266-267},
year = {1979},
doi = { https://doi.org/10.1137/1021044}
}

@book{0078096,
  author = {LeVeque, Randall J.},
  isbn = {978-3-7643-2723-1},
  pages = {1-214},
  publisher = {Birkhäuser},
  series = {Lectures in mathematics},
  title = {Numerical {M}ethods for {C}onservation {L}aws},
  edition= {Second},
  year = 1992
}

@Article{Eckart1936,
author={Eckart, Carl
and Young, Gale},
title={The approximation of one matrix by another of lower rank},
journal={Psychometrika},
year={1936},
month={Sep},
day={01},
volume={1},
number={3},
pages={211-218},
issn={1860-0980},
doi={https://doi.org/10.1007/BF02288367},
}

@Inbook{Houska2012,
author="Houska, Boris
and Logist, Filip
and Diehl, Moritz
and Van Impe, Jan",
title="A tutorial on numerical methods for state and parameter estimation in nonlinear dynamic systems",
bookTitle="Identification for Automotive Systems",
chapter="5",
year="2012",
publisher="Springer London",
address="London",
pages="67--88",
isbn="978-1-4471-2221-0",
doi="https://doi.org/10.1007/978-1-4471-2221-0_5"
}

@article{Gunning,
author = {Gunning, David and Aha, David W.},
title = {{DARPA}'s Explainable Artificial Intelligence Program},
journal = {AI Magazine},
volume = {40},
number = {2},
pages = {44-58},
doi = {https://doi.org/10.1609/aimag.v40i2.2850},
year = {2019}
}

@article{National,
author = {{National Academy of Engineering and National Academies of Sciences, Engineering,
and Medicine}},
title = {Foundational Research Gaps and Future Directions for Digital Twins},
journal = {The National
Academies Press, Washington, DC},
doi = {https://doi.org/10.17226/26894},
year = {2024}}

@article{Steven2015,
title = {Compressed sensing and dynamic mode decomposition},
journal = {Journal of Computational Dynamics},
volume = {2},
number = {2},
pages = {165-191},
year = {2015},
issn = {2158-2491},
doi = {https://doi.org/10.3934/jcd.2015002},
author = {Steven L. Brunton and Joshua L.  Proctor and Jonathan H.  Tu and J. Nathan Kutz}
}

@article{annurev015835,
   author = "Schmid, Peter J.",
   title = "Dynamic Mode Decomposition and Its Variants", 
   journal= "Annual Review of Fluid Mechanics",
   year = "2022",
   volume = "54",
   number = "Volume 54, 2022",
   pages = "225-254",
   doi = "https://doi.org/10.1146/annurev-fluid-030121-015835",
   publisher = "Annual Reviews",
   issn = "1545-4479"
}

@article{doi130914619,
author = {Ionita, A. C. and Antoulas, A. C.},
title = {Data-Driven Parametrized Model Reduction in the Loewner Framework},
journal = {SIAM Journal on Scientific Computing},
volume = {36},
number = {3},
pages = {A984-A1007},
year = {2014},
doi = {https://doi.org/10.1137/130914619}
}

@article{Franz16032014,
author = {T. Franz and R. Zimmermann and S. Görtz and N. Karcher},
title = {Interpolation-based reduced-order modelling for steady transonic flows via manifold learning},
journal = {International Journal of Computational Fluid Dynamics},
volume = {28},
number = {3-4},
pages = {106--121},
year = {2014},
publisher = {IAHR Website},
doi = {https://doi.org/10.1080/10618562.2014.918695}
}

@article{doiJ065798,
author = {Zastrow, Benjamin G. and Chaudhuri, Anirban and Willcox, Karen E. and Ashley, Anthony and Henson, Michael Chamberlain},
title = {Block-Structured Operator Inference for Coupled Multiphysics Model Reduction},
journal = {AIAA Journal},
volume = {0},
number = {0},
pages = {1-16},
year = {0},
doi = {https://doi.org/10.2514/1.J065798}
}

@article{doi22M1481658,
author = {Andreuzzi, Francesco and Demo, Nicola and Rozza, Gianluigi},
title = {A Dynamic Mode Decomposition Extension for the Forecasting of Parametric Dynamical Systems},
journal = {SIAM Journal on Applied Dynamical Systems},
volume = {22},
number = {3},
pages = {2432-2458},
year = {2023},
doi = {https://doi.org/10.1137/22M1481658
}}

@article{CICCI20231,
title = {Uncertainty quantification for nonlinear solid mechanics using reduced order models with Gaussian process regression},
journal = {Computers \& Mathematics with Applications},
volume = {149},
pages = {1-23},
year = {2023},
issn = {0898-1221},
doi = {https://doi.org/10.1016/j.camwa.2023.08.016},
author = {Ludovica Cicci and Stefania Fresca and Mengwu Guo and Andrea Manzoni and Paolo Zunino}}

@article{GUO2018807,
title = {Reduced order modeling for nonlinear structural analysis using Gaussian process regression},
journal = {Computer Methods in Applied Mechanics and Engineering},
volume = {341},
pages = {807-826},
year = {2018},
issn = {0045-7825},
doi = {https://doi.org/10.1016/j.cma.2018.07.017},
author = {Mengwu Guo and Jan S. Hesthaven}
}

@article{articleManzoni,
author = {Franco, Nicola Rares and Manzoni, Andrea and Zunino, Paolo},
year = {2023},
month = {3},
pages = {483–524},
title = {A deep learning approach to Reduced Order Modelling of parameter dependent partial differential equations},
volume = {92},
journal = {Mathematics of Computation},
doi = {https://doi.org/10.1090/mcom/3781}
}

@article{LOAN200085,
title = {The ubiquitous Kronecker product},
journal = {Journal of Computational and Applied Mathematics},
volume = {123},
number = {1},
pages = {85-100},
year = {2000},
note = {Numerical Analysis 2000. Vol. III: Linear Algebra},
issn = {0377-0427},
doi = {https://doi.org/10.1016/S0377-0427(00)00393-9},
url = {https://www.sciencedirect.com/science/article/pii/S0377042700003939},
author = {Charles F.Van Loan}}

@article{104913868,
    author = {Sayadi, Taraneh and Schmid, Peter J. and Richecoeur, Franck and Durox, Daniel},
    title = {Parametrized data-driven decomposition for bifurcation analysis, with application to thermo-acoustically unstable systems},
    journal = {Physics of Fluids},
    volume = {27},
    number = {3},
    pages = {037102},
    year = {2015},
    month = {03},
    issn = {1070-6631},
    doi = {https://doi.org/10.1063/1.4913868},
}

@article{mcquarrie2025bayesian,
  title={Bayesian learning with {G}aussian processes for low-dimensional representations of time-dependent nonlinear systems},
  author={McQuarrie, Shane A and Chaudhuri, Anirban and Willcox, Karen E and Guo, Mengwu},
  journal={Physica D: Nonlinear Phenomena},
  volume={475},
  pages={134572},
  year={2025},
  publisher={Elsevier}
}

@article{fresca2022pod,
  title={{POD-DL-ROM}: {E}nhancing deep learning-based reduced order models for nonlinear parametrized PDEs by proper orthogonal decomposition},
  author={Fresca, Stefania and Manzoni, Andrea},
  journal={Computer Methods in Applied Mechanics and Engineering},
  volume={388},
  pages={114181},
  year={2022},
  publisher={Elsevier}
}

@article{sapienza2024differentiable,
  title={Differentiable programming for differential equations: {A} review},
  author={Sapienza, Facundo and Bolibar, Jordi and Sch{\"a}fer, Frank and Groenke, Brian and Pal, Avik and Boussange, Victor and Heimbach, Patrick and Hooker, Giles and P{\'e}rez, Fernando and Persson, Per-Olof and others},
  journal={arXiv preprint arXiv:2406.09699},
  year={2024}
}

\end{document}